\documentclass[aip,amsmath,amssymb,numerical,reprint,%
  longbibliography]{revtex4-2}

\usepackage{dcolumn}
\usepackage{bm}

\usepackage[utf8]{inputenc}
\usepackage[T1]{fontenc}
\usepackage{mathptmx}
\usepackage{etoolbox}

\usepackage{mathtools}
\usepackage{MnSymbol}

\newcommand{\be}{\begin{equation}}
\newcommand{\ee}{\end{equation}}
\newcommand{\bea}{\begin{eqnarray}}
\newcommand{\eea}{\end{eqnarray}}

\def\softd{{\leavevmode\setbox1=\hbox{d}%
          \hbox to 1.05\wd1{d\kern-0.4ex{\char039}\hss}}}

\newcommand{\MariasFirstName}{M\' aria}
\newcommand{\MariasLastName}{Luk\'a\v{c}ov\'a-Medvi\softd ov\'a}

\makeatletter
\def\@email#1#2{%
 \endgroup
 \patchcmd{\titleblock@produce}
  {\frontmatter@RRAPformat}
  {\frontmatter@RRAPformat{\produce@RRAP{*#1\href{mailto:#2}{#2}}}%
    \frontmatter@RRAPformat}
  {}{}
}%
\makeatother

\begin{document}

\title{Fundamentals of the Oldroyd-B model revisited: Tensorial
  vs. vectorial theory}

\author{Aaron Brunk}
\email[]{abrunk@uni-mainz.de}
\affiliation{Institute of Mathematics, Johannes Gutenberg
  University Mainz, Staudingerweg 9, 55128 Mainz, Germany}

\author{Joydip Chaudhuri}
\email[]{chaudhurij@mpip-mainz.mpg.de}
\affiliation{Max Planck Institute for Polymer Research,
  Ackermannweg 10, 55128 Mainz, Germany}

\author{{\MariasFirstName} {\MariasLastName}}
\email[]{lukacova@mathematik.uni-mainz.de}
\affiliation{Institute of Mathematics, Johannes Gutenberg
  University Mainz, Staudingerweg 9, 55128 Mainz, Germany}

\author{Burkhard D\"{u}nweg}
\email[]{duenweg@mpip-mainz.mpg.de}
\affiliation{Max Planck Institute for Polymer Research,
  Ackermannweg 10, 55128 Mainz, Germany}
\affiliation{Department of Chemical and Biological
  Engineering, Monash University, Clayton, Victoria 3800, Australia}

\date{\today}

\begin{abstract}
  The standard derivation of the Oldroyd-B model starts from a coupled
  system of the momentum equation for the macroscopic flow on the one
  hand, and Fokker-Planck dynamics for molecular dumbbells on the
  other. The constitutive equation is then derived via a closure based
  upon the second moment of the end-to-end vector distribution. We
  here present an alternative closure that is rather based upon the
  first moment, and gives rise to an even simpler constitutive
  equation. We establish that both closures are physically sound,
  since both can be derived from (different) well-defined
  non-equilibrium ensembles, and both are consistent with the Second
  Law of thermodynamics. In contrast to the standard model, the new
  model has a free energy and a dissipation rate that are both regular
  at vanishing conformation tensor. We speculate that this might
  perhaps alleviate the well-known high Weissenberg number problem,
  i.~e. severe numerical instabilities of the standard model at large
  flow rates. As the new model permits a trivial solution (vanishing
  conformation tensor, vanishing polymer stress), an extension may be
  needed, which includes Langevin noise in order to model thermal
  fluctuations.
\end{abstract}

\maketitle

\section{Introduction}
\label{sec:Intro}

The Oldroyd-B model~\cite{birdDynamicsPolymericLiquids1987a,
  larsonStructureRheologyComplex1999, owensComputationalRheology2002,
  phan-thienUnderstandingViscoelasticityBasics2013,
  OldroydBModel2021Wikipedia} is one of the most popular models to
describe the rheology of polymer solutions. Except for the
incompressibility condition for the flow field, and the momentum
conservation equation, it features a constitutive equation for the
stress, which is an additional time-dependent equation of motion in
order to take into account viscoelasticity, or, in other words, the
fact that intramolecular relaxation happens on a time scale that is
comparable to the relevant time scales of the flow.

Despite its popularity, the model is known to run into severe
stability problems when solving the equations numerically and studying
Weissenberg numbers that are too large --- recall that the
dimensionless Weissenberg number is defined as the product of shear
rate and molecular relaxation time. This well-known ``high Weissenberg
number problem'' (HWNP) has been confirmed and discussed in numerous
computer simulations, e.~g. (to name just one out of many) in
Ref.~\onlinecite{bajajCoilstretchTransitionBreakdown2008}. The
problem must still be considered as essentially unsolved.

The standard derivation of the
model~\cite{birdDynamicsPolymericLiquids1987a} starts from a two-scale
description: On the one hand, the flow on the macroscale is described
by the Navier-Stokes equation, augmented by an additional polymeric
stress. On the other hand, the polymer component is sketched by an
extremely simple model --- a set of non-interacting Hookean dumbbells
whose dynamics is described by a simple Langevin equation (or
equivalent Fokker-Planck equation, FPE) which describes the balance
between spring forces, friction forces and stochastic forces. One then
observes that the Kramers (or virial) expression for the microscopic
polymer-induced stress is directly related to the so-called
conformation tensor, i.~e. the tensor product of the dumbbell's
end-to-end vector with itself. Then, without deep justification, one
simply averages the tensor with the time-dependent non-equilibrium
distribution function for the end-to-end vector, i.~e. the solution of
the FPE. On the one hand, this provides a prescription how to obtain
the polymer stress on the macroscale from the microscopic
configurations and the microscale dynamics. On the other hand, the
Fokker-Planck dynamics gives rise to an equation of motion for the
such-defined average conformation tensor. Because of the simplicity of
the microscopic model, this equation happens to be closed, and is
nothing but the constitutive equation of the standard Oldroyd-B model.

For good reasons, this procedure of derivation has been termed
``closure'', in order to express the fact that there is a certain
amount of arbitrariness and approximation that underlies the
reasoning. In the present paper, we wish to point out that the
standard closure corresponds to a \emph{specific choice} of the
underlying statistical-mechanical non-equilibrium ensemble, which is
by no means the only choice that is possible. Indeed, from the
Mori-Zwanzig
formalism~\cite{forsterHydrodynamicFluctuationsBroken1995,
  j.phansenTheorySimpleLiquids2013} we know that the very first step
to derive simplified dynamic equations is the identification (or,
actually, \emph{choice}) of so-called ``slow variables'',
i.~e. observables whose dynamics is assumed to be so slow that they
should be taken into account as degrees of freedom on the
macroscale. Therefore, it is only the remaining ``fast'' variables
that one should average over. The common derivation of the standard
closure does not clearly make that distinction, and this is what makes
it conceptually slightly obscure. Note that the identification of slow
variables, combined with averaging over the fast ones, may be viewed
as the definition of a non-equilibrium \emph{ensemble}. As a matter of
fact, we will demonstrate that the standard Oldroyd-B equation may be
derived within that conceptual framework by simply postulating that
the slow variable is actually the conformation tensor as
such. However, other ensembles may be defined as well, simply by
picking other slow variables. In the present paper, we wish to analyze
the consequences of rather taking the end-to-vector as the
ensemble-defining observable, and to demonstrate that this is a valid
choice, leading to a new fully consistent theory.  There are further
differences concerning the number of dumbbells that contribute to the
observable; these details will become clear below. The modified
ensemble then gives rise to a modified constitutive equation, and is
therefore an alternative closure, which we call the ``modified
Oldroyd-B model''. It can be verified that both closures give rise to
a macroscopic model that satisfies the Second Law of thermodynamics;
however, the free energy is different for the two closures (which is
not surprising, given that the ensembles differ). From a practical
point of view, it turns out that the modified model is based not on
the \emph{second} moment of the end-to-end vector distribution
function (as is the standard version), but rather on its \emph{first}.

There seems to be widespread belief in the community that basing the
theory on the first moment is fundamentally impossible, because it is
believed to vanish identically, for symmetry reasons. The argument is
simply that a dumbbell is intrinsically symmetric, such that a simple
re-labeling of the two beads would turn the dumbbell's end-to-end
vector $\vec{q}$ into $- \vec{q}$. However, nothing prevents us from
just picking one of the two orientations and then basing the
description on this. The symmetry only means that we may as well just
pick the other orientation, and nevertheless obtain completely
identical results. Indeed, let us assume that we have a small volume
element which contains a number of dumbbells. The typical model
assumption is that the element is (i) small enough, such that the flow
within it may be considered as homogeneous (with constant velocity
gradient tensor), but also (ii) large enough, such that the
contributing dumbbells may be described in terms of a probability
density for the end-to-end vector, $P(\vec{q})$. Typically, the
nonzero velocity gradient will result in some local alignment of
dumbbells, and we now wish to map that configuration (say, at some
initial time $t=0$) onto a probability density $P(\vec{q})$.  Now,
going through the molecules and picking their orientation at random
will indeed produce a density with vanishing first moment, and that
property will then continue to be present in the dynamics. However, we
can also adopt a different convention: We pick some (arbitarily
chosen) orientation for one particular molecule and then choose the
orientation of the other molecules such that they are maximally
aligned with the original one. It is important to realize that this
has to be done not only within the selected volume element, but also
in all the other ones, whose orientations have to be mutually
consistent in order to make sure that $\vec{q}$ can, in the continuum
limit, be described as a smooth vector field. It is clear that with
such a labeling convention we will get a $P(\vec{q})$ which does, at
$t=0$, exhibit a nonzero first moment. As the dynamics of $P$ is
assumed to be smooth with respect to time, the nonzero first moment
will then also exist for some later times. It should be noted that
this argument of course only works in a non-equilibrium
(shear-aligned) situation --- in strict thermal equilibrium the first
moment is of course indeed always zero, simply because in this
situation the attempt to pick the orientations in a mutually aligned
fashion will be unsuccessful. The whole discussion of kinetic theory
later in the paper should therefore be read with the notion in mind
that we do admit a nonzero first moment, by imposing a suitable
initial condition, which in turn corresponds to a specific labeling
convention of the molecules.

An interesting feature of the modified model is that it removes the
singularity of the free energy at vanishing conformation tensor, which
is present for the standard version. This singularity also occurs in
the dissipation rate, indicative of the system's ability to absorb
arbitrarily large amounts of energy. The well-known
HWNP~\cite{bajajCoilstretchTransitionBreakdown2008} may perhaps be
related to this; therefore one may speculate that the new model might
perhaps have a reduced or even non-existent HWNP. For this reason, we
think it is worthwhile to subject the new model to thorough numerical
tests; this is however deferred to future work.

It should be mentioned that we originally realized the ensemble
problems, and the ambiguity of the closure, not by the reasoning
presented here. Rather, we had set out to derive hydrodynamic
equations for viscoelastic phase separation, i.~e. the spinodal
decomposition of viscoelastic fluids. We did this with a completely
different theoretical approach: Starting from essentially the same
phantom dumbbell picture that underlies the derivation of the
Oldroyd-B model, we subjected the system to coarse-graining, which
resulted in a field-theoretic Hamiltonian and a field-theoretic
dissipation rate. Combining this with the GENERIC
formalism~\cite{grmelaDynamicsThermodynamicsComplex1997,
  ottingerDynamicsThermodynamicsComplex1997,
  ottingerEquilibriumThermodynamics2005}, and various simplifying
approximations, we arrived at a set of equations, which, when assuming
a homogeneous system that does not unmix, reduces exactly to the
Oldroyd-B system --- however, \emph{not} in the standard version but
the modified one.  We then realized that the difference was due to a
difference in the underlying non-equilibrium ensemble, and made some
remarks along this line in our publication~\cite{
  spillerSystematicDerivationHydrodynamic2021a}, which describes the
derivation in detail. However, a detailed and careful analysis of the
issue, and in particular a systematic comparison of the two variants,
was missing. The present paper is intended to provide these missing
items, in order to put the new variant on a sound theoretical
basis. We believe these considerations are relevant not only within
the framework of the Oldroyd-B model, but for rheology in general.

We will start out from the standard formulation of the Oldroyd-B
model, and then generalize it to also include the modified
version. This generalized model is then subjected to theoretical
scrutiny, where we discuss the conservation of the conformation
tensor's positivity and establish the validity of the Second Law. We
only then discuss the underlying statistical mechanics, and first
outline the general formalism of a non-equilibrium ensemble based upon
the notion of slow observables. This allows us to construct the free
energies of both variants from the partition function, and also their
constitutive equations from Fokker-Planck theory, for which we show
the Second Law as well. We will conclude with a brief summary.

\section{Basic formulation}
\label{sec:Formulation}

Our starting point are the equations of motion of the Oldroyd-B model.
Denoting the fluid flow velocity field with $\vec{v} (\vec{r}, t)$,
where $\vec{r}$ is the spatial coordinate and $t$ denotes time, we may
introduce the operator for the convective derivative via
\be
D_t = 
\frac{\partial}{\partial t} + \vec{v} \cdot \nabla
 = \partial_t + \vec{v} \cdot \nabla ,
\ee
or, in terms of Cartesian components $\alpha, \beta, \ldots$,
for which we assume the Einstein summation convention,
\be
D_t = \partial_t + v_\alpha \frac{\partial}{\partial r_\alpha}
= \partial_t + v_\alpha \partial_\alpha .
\ee
The incompressibility condition for the flow is written as
\be
\nabla \cdot \vec{v} = 0 ,
\ee
and the momentum conservation equation reads
\be
\rho D_t \vec{v} = - \nabla p + \nabla \cdot \tensor{T} ,
\ee
or, in components,
\be
\rho D_t v_\alpha = - \partial_\alpha p
+ \partial_\beta T_{\alpha \beta} .
\ee
Here $\rho$ is the constant mass density of the fluid, $p$ is the
scalar pressure (which should be viewed as a Lagrange multiplier or
``reaction stress'' enforcing incompressibility), and $\tensor{T}$ the
viscoelastic stress tensor.

As $\tensor{T}$ is symmetric, it does not matter if the derivative in
$\nabla \cdot \tensor{T}$ acts on the first or the second
index. Introducing the velocity gradient tensor $\tensor{\kappa}$ via
\be
\kappa_{\alpha \beta} = \partial_\beta v_\alpha ,
\ee
we can define the upper convected derivative of a symmetric
second-rank tensor as
\be
\delta_t \tensor{T}
= D_t \tensor{T}
- \tensor{\kappa} \cdot \tensor{T}
- \tensor{T} \cdot \tensor{\kappa}^T .
\ee
Here the dot product is to be understood as a conventional
matrix product, i.~e., in components the definition reads
\bea
\nonumber
\delta_t T_{\alpha \beta}
& = & D_t T_{\alpha \beta}
- \kappa_{\alpha \gamma} T_{\gamma \beta}
- T_{\alpha \gamma} \kappa_{\beta \gamma}
\\
& = &
D_t T_{\alpha \beta}
- T_{\beta \gamma} \partial_\gamma v_\alpha
- T_{\alpha \gamma} \partial_\gamma v_\beta .
\eea
It should be noted that from a formal point of view one may also
define other convected derivatives (for example, the so-called ``lower
convected derivative''~\cite{birdDynamicsPolymericLiquids1987});
however, in the standard Oldroyd-B model the upper convected
derivative occurs, as a direct consequence of its kinetic-theory
derivation.

Furthermore, we can define the deformation rate tensor
\be
\tensor D = \frac{1}{2}
\left( \tensor{\kappa} + \tensor{\kappa}^T \right) .
\ee
In particular, we find for the upper convected derivative of the unit
tensor
\be
\delta_t \tensor{1} = - 2 \tensor{D} .
\ee
Finally, we introduce the solvent viscosity $\eta_s > 0$, the polymer
viscosity $\eta_p \ge 0$, the total viscosity
$\eta = \eta_s + \eta_p$, a molecular relaxation time $\lambda > 0$, an
associated modulus $G = \eta_p / \lambda$, and the dimensionless ratio
\be
\Gamma = \frac{\eta_p}{\eta_s} \ge 0 ,
\ee
which may be viewed as a parameter which tells us how strong
viscoelastic effects are expected to be. Note that these parameters
are all macroscopic, with no direct reference to a molecular picture,
with the exception of $\lambda$, which denotes the typical time scale
for molecular processes (without specifying details about these
processes).

With these notational preliminaries, which may be used for essentially
any rheological model for polymer solutions, we can now focus on the
specific formulation of the Oldroyd-B model. According to
Refs.~\onlinecite{birdDynamicsPolymericLiquids1987a} or
\onlinecite{OldroydBModel2021Wikipedia}, its constitutive equation can
be written in the following compact form:
\be
\label{eq:OldroydB1}
\tensor{T} + \lambda \delta_t \tensor{T} =
2 \eta \tensor{D} + 2 \eta_s \lambda \delta_t \tensor{D} .
\ee
For the further development it is useful to slightly re-write the
equations. Firstly, we recall the standard constitutive relation for
the solvent stress $\tensor{\tau}^{(s)}$, which is that of a Newtonian
fluid:
\be
\tensor{\tau}^{(s)} = 2 \eta_s \tensor{D} ,
\ee
such that, due to incompressibility,
\be
\nabla \cdot \tensor{\tau}^{(s)} = \eta_s \nabla^2 \vec{v} .
\ee
We can thus transform the right-hand side (RHS) of
Eq.~\ref{eq:OldroydB1} as follows:
\bea
\nonumber
& &
2 \eta \tensor{D} + 2 \eta_s \lambda \delta_t \tensor{D}
\\
\nonumber
& = &
2 \eta_s
\left( \tensor{D} + \lambda \delta_t \tensor{D} \right)
+
2 \eta_p \tensor{D}
\\
\nonumber
& = &
\tensor{\tau}^{(s)} + \lambda \delta_t \tensor{\tau}^{(s)}
-
\eta_p \delta_t \tensor{1}
\\
& = &
\tensor{\tau}^{(s)} + \lambda \delta_t \tensor{\tau}^{(s)}
-
G \lambda \delta_t \tensor{1} .
\eea
As the total stress $\tensor{T}$ is the sum of the polymer stress
$\tensor{\tau}^{(p)}$ and the solvent stress $\tensor{\tau}^{(s)}$,
we may thus subtract the latter contribution from Eq.~\ref{eq:OldroydB1}
and obtain
\be
\tensor{\tau}^{(p)} + \lambda \delta_t \tensor{\tau}^{(p)}
=
-
G \lambda \delta_t \tensor{1} .
\ee
We now decompose
\be
\tensor{\tau}^{(p)} = \tensor{\sigma} - G \tensor{1}
\ee
and obtain
\be
\tensor{\sigma} - G \tensor{1} + \lambda \delta_t \tensor{\sigma} = 0
\ee
or
\be
\label{eq:OldroydB2}
\delta_t \tensor{\sigma}
= - \frac{1}{\lambda} \left( \tensor{\sigma}
- G \tensor{1} \right) ,
\ee
which is thus just a transformed way of writing the Oldroyd-B
constitutive equation. In terms of the transformed variables,
the momentum equation reads
\be
\rho D_t \vec{v} = - \nabla p + \eta_s \nabla^2 \vec{v}
+ \nabla \cdot \tensor{\sigma} .
\ee
In this form, the equations are more amenable to theoretical analysis.

We now introduce a dimensionless tensor $\tensor{C}$, commonly referred
to as ``conformation tensor'', via
\be
\tensor{\sigma} = G \tensor{C} =
\Gamma \frac{\eta_s}{\lambda} \tensor{C} .
\ee
With this re-parametrization, the constitutive equation reads
\be
\label{eq:OldroydB3}
\delta_t \tensor{C}
= - \frac{1}{\lambda} \left( \tensor{C} - \tensor{1} \right) ,
\ee
while the momentum equation is transformed to
\be
\rho D_t \vec{v} = - \nabla p + \eta_s \nabla^2 \vec{v}
+ \Gamma \frac{\eta_s}{\lambda} \nabla \cdot \tensor{C} .
\ee
Since the parameters $\rho$, $\eta_s$, and $\lambda$ have physical
dimensions that are mutually different, we may set them to unity
in order to define a natural unit system for the problem. We
thus obtain for the non-dimensionalized set of Oldroyd-B equations:
\bea
\label{eq:OldroydB4}
\nabla \cdot \vec{v} & = & 0 , \\
\label{eq:OldroydB5}
D_t \vec{v} + \nabla p - \Gamma \nabla \cdot \tensor{C}
& = & \nabla^2 \vec{v} , \\
\label{eq:OldroydB6}
\delta_t \tensor{C} & = & - \tensor{C} + \tensor{1} .
\eea
In the last two equations, we have grouped the contributions in such a
way that all the terms on the left-hand side (LHS) exhibit the same
behavior under time reversal, while the terms on the RHS exhibit the
opposite behavior. This allows us to clearly identify the terms on the
RHS as the \emph{dissipative} contributions, while those on the LHS
are the time derivatives and the \emph{conservative}
contributions. This necessary split-up, which comes directly from
time-reversal symmetry, is central for the further development.

In what follows, we will discuss a \emph{generalized} Oldroyd-B
model. We wish to keep the momentum equation as-is, because it
involves little more than the mere definition of stress. Similarly, we
wish to leave the LHS of Eq.~\ref{eq:OldroydB6} unchanged --- the
description of the kinematics of $\tensor{C}$ in terms of the upper
convected derivative is widely accepted in the rheology community, and
shall not be challenged in the present paper either --- even less so,
as it is a clear and unrefutable consequence of the kinetic theory
outlined below. However, the dissipative part of
Eq.~\ref{eq:OldroydB6} shall be generalized to
\be
\label{eq:OldroydB7}
\delta_t \tensor{C} = - \tensor{C} + \alpha_d \tensor{1} ,
\ee
where we introduce a dimensionless scalar parameter $\alpha_d$ (the
subscript $d$ may be read as ``dynamics'' or ``dissipation''). In
other words, we generalize the dissipative part of the constitutive
equation to a first-order polynomial in $\tensor{C}$, where the
pre\-factor of the linear term is fixed by the requirement of
relaxational dynamics, plus the chosen unit system. As we do not wish
to discuss models in which the dissipative terms depend non-linearly
on $\tensor{C}$, this is the most general form.

\section{Positivity of the conformation tensor}
\label{sec:positivity}

It will become clear later that positive (semi-)definiteness of
$\tensor{C}$ plays a crucial role in the theory. Let us therefore
assume that at time $t = 0$ we start out with a tensor that is
positive (semi-)definite (throughout the volume of the system), and
investigate under which conditions this property is conserved under
the dynamics.

At $t = 0$ the spectral decomposition of $\tensor{C}$ reads
\be
\label{eq:SpectralDecomposition}
\tensor{C} (t=0) = \sum_k \lambda_k \vec{n}^{(k)} \vec{n}^{(k)T} ,
\ee
where $\lambda_k > 0$ are the positive eigenvalues, and
$\vec{n}^{(k)}$ the corresponding normalized eigenvectors,
$\vec{n}^{(k)} \cdot \vec{n}^{(k)} = 1$. This means we can also write
\be
\label{eq:DecompositionOFCStart}
\tensor{C} (t=0) = \sum_k \vec{p}^{(k)} (t=0)\vec{p}^{(k)T} (t=0)
\ee
with $\vec{p}^{(k)} (t=0) = \lambda_k^{1/2} \vec{n}^{(k)}$.

We therefore try the ansatz
\be
\label{eq:DecompositionOFC}
\tensor{C} (t) = \sum_k \vec{p}^{(k)} (t)\vec{p}^{(k)T} (t) ,
\ee
which is manifestly positive semi-definite, and find
\bea
\nonumber
\delta_t \tensor{C} + \tensor{C}
& = &
\sum_k \left[
  \left( D_t \vec{p}^{(k)} \right) \vec{p}^{(k)T}
  + \vec{p}^{(k)} \left( D_t \vec{p}^{(k)T} \right)
  \right.
  \\
  \nonumber
  &&
  \left.
  - \left( \tensor{\kappa} \cdot \vec{p}^{(k)} \right) \vec{p}^{(k)T}
  - \vec{p}^{(k)} \left( \tensor{\kappa} \cdot \vec{p}^{(k)} \right)^T
  + \vec{p}^{(k)} \vec{p}^{(k)T}
\right]
\\
\nonumber
& = &
\sum_k \left[   
  \left( D_t \vec{p}^{(k)} - \tensor{\kappa} \cdot \vec{p}^{(k)}
    + \frac{1}{2} \vec{p}^{(k)} \right) \vec{p}^{(k)T}
  \right.
  \\
  &&
  \left.
  + \vec{p}^{(k)}
  \left( D_t \vec{p}^{(k)}
    - \tensor{\kappa} \cdot \vec{p}^{(k)}
    + \frac{1}{2} \vec{p}^{(k)} \right)^T
\right] .
\label{eq:TensorDynamicsVsVectorDynamics}
\eea
This, however, means that if we propagate the vectors
$\vec{p}^{(k)}$ according to
\be
\label{eq:VectorialEquationOfMotion}
D_t \vec{p}^{(k)} - \tensor{\kappa} \cdot \vec{p}^{(k)}
+ \frac{1}{2} \vec{p}^{(k)} = 0 ,
\ee
then Eq.~\ref{eq:DecompositionOFC} will solve
\be
\label{eq:HomogeneousOldroydB}
\delta_t \tensor{C} + \tensor{C} = 0 .
\ee

We thus see that for $\alpha_d = 0$ the constitutive equation
conserves the property of positive semi-definiteness. The question if
in this case also strict positive-definiteness is conserved, such that
all eigenvalues, if assumed to be positive at $t=0$, will remain
positive throughout the dynamics, is more subtle. We note that for
strict positive-definiteness we need that the vectors $\vec{p}^{(k)}$
form a basis, meaning that their number must equal the spatial
dimension, that they must be nonzero, and that they are linearly
independent. It should be noted that we need only linear independence
and not mutual orthogonality. As a matter of fact, one should expect
that the dynamics will typically not maintain mutual orthogonality,
even if it holds at $t=0$. It is also reasonable to assume (although
not straightforward to prove) that linear independence will remain
intact during the dynamics. If also each of the vectors never goes
through zero, then indeed strict positive definiteness is maintained
--- but this is far from obvious for arbitrary flows. Fortunately,
however, it will become clear below that for $\alpha_d = 0$ positive
semi-definiteness is sufficient for the model to be sound.

For general $\alpha_d$, we replace the true dynamics by a discretized
and thus approximate dynamics that is inspired by numerical
time-stepping schemes. We introduce a small time step $h$ and
propagate the system for the duration of one time step as follows: (i)
Update the dynamical variables by solving the system
\bea
\label{eq:Strang1}
\nabla \cdot \vec{v} & = & 0 , \\
\label{eq:Strang2}
D_t \vec{v} + \nabla p - \Gamma \nabla \cdot \tensor{C}
& = & \nabla^2 \vec{v} , \\
\label{eq:Strang3}
\delta_t \tensor{C} + \tensor{C}
& = & 0 ,
\eea
for the duration of $h/2$; (ii) update them according to
\bea
\label{eq:Strang4}
\partial_t \vec{v} & = & 0 , \\
\label{eq:Strang5}
\partial_t \tensor{C} & = & \alpha_d \tensor{1} ,
\eea
for the duration of $h$; and (iii) update them for $h/2$, again
following the prescription given in step (i). This scheme is known
as Strang splitting, which yields the correct dynamics up to errors
of order $h^2$.

Now, it is clear from the previous analysis that steps (i) and (iii)
do not alter the positive-semidefinite character of $\tensor{C}$,
provided that the initial value at the beginning of the update is
positive-semidefinite. Step (ii) will also maintain positive
semi-definiteness, provided
\be
\alpha_d \ge 0 ,
\ee
which we will from now on postulate as a necessary condition for a
valid generalized Oldroyd-B model. As a matter of fact, for $\alpha_d
> 0$ the eigenvalues are increased during step (ii), and thus in this
case the dynamics even conserves positive-definiteness. This is in
accord with known results from the
literature~\cite{hulsenSufficientConditionPositive1990}.

\section{Hamiltonian I: Analysis of conservative dynamics}
\label{sec:ConservativeDynamics}

For the conservative part of the dynamics we have the equations
of motion
\bea
\nabla \cdot \vec{v} & = & 0 , \\
D_t \vec{v} + \nabla p - \Gamma \nabla \cdot \tensor{C}
& = & 0 , \\
\delta_t \tensor{C} & = & 0 .
\eea
Per construction, these equations are time-reversal
symmetric. However, for being conservative, the dynamics must also
conserve the underlying Hamiltonian ${\cal H}$, which should, in the
present isothermal setting, be interpreted as the Helmholtz free
energy of the system. As the model is a local field theory, the
Hamiltonian must be written as a functional of the fields $\vec{v}$
and $\tensor{C}$. As the kinetic energy density is $\vec{v}^2/2$, the
only reasonable functional that is consistent with the requirement of
a \emph{local} free energy density has the form (in $d$ spatial
dimensions)
\be
{\cal H} = \int d \vec{r} \,
\left[ \frac{\vec{v}^2}{2} + f (\tensor{C}) \right] ,
\ee
where $f$ is a scalar function of the conformation tensor. It should
be noted that for any volume integral we assume either a bounded
domain with periodic boundary conditions, or an infinite domain with
rapidly decaying fields, such that an integration by parts will not
involve any boundary terms. Clearly, $f$ must be chosen in such a way
that the Hamiltonian is conserved, i.~e.
\be
\frac{d {\cal H}}{dt} = 0 .
\ee
To this end, we first consider the equation of motion
\be
\label{eq:equationofmotion}
\partial_t v_\alpha = - v_\beta \partial_\beta v_\alpha
- \partial_\alpha p + \Gamma \partial_\beta C_{\alpha \beta} .
\ee
Multiplying with $v_\alpha$ and integrating over space yields, after
some integration by parts, and making use of incompressibility,
\be
\label{eq:eqofmotionHamiltonian}
\frac{d}{dt} \int d \vec{r} \, \frac{\vec{v}^2}{2} =
- \Gamma \int d \vec{r} \,
\kappa_{\alpha \beta} C_{\alpha \beta} .
\ee
Similarly, we study
\be
\partial_t C_{\alpha \beta} =
- v_\gamma \partial_\gamma C_{\alpha \beta}
+ C_{\beta \gamma} \kappa_{\alpha \gamma}
+ C_{\alpha \gamma} \kappa_{\beta \gamma} .
\ee
Multiplying with $\chi_{\alpha \beta} := \partial f / \partial
C_{\alpha \beta}$, followed by integration over space, yields, again
after some integration by parts, and index re-labeling, making
use of the symmetry of $\tensor{C}$ and $\tensor{\chi}$,
\bea
\nonumber
&&
\frac{d}{dt} \int d \vec{r} \, f
\\
\nonumber
& = &
\int d \vec{r} \, \left[
-  v_\gamma \chi_{\alpha \beta} \partial_\gamma C_{\alpha \beta}
+ \chi_{\alpha \beta} C_{\beta \gamma} \kappa_{\alpha \gamma}
+ \chi_{\alpha \beta} C_{\alpha \gamma} \kappa_{\beta \gamma}
\right]
\\
\nonumber
& = &
- \int d \vec{r} \, v_\gamma \partial_\gamma f
+ 2 \int d \vec{r} \, \kappa_{\alpha \beta}
C_{\beta \gamma} \chi_{\gamma \alpha}
\\
& = &
2 \int d \vec{r} \, \kappa_{\alpha \beta}
C_{\beta \gamma} \chi_{\gamma \alpha} .
\eea
Therefore
\be
\label{eq:ConditionConservation}
0 = \frac{d {\cal H}}{dt} =
\int d \vec{r} \, \kappa_{\alpha \beta}
\left[ 2 C_{\beta \gamma} \chi_{\gamma \alpha}
  - \Gamma C_{\alpha \beta} \right] .
\ee
Now, the conservation of ${\cal H}$ must hold for any
velocity gradient tensor. An obvious solution to this
problem is therefore the requirement
\be
2 C_{\beta \gamma} \chi_{\gamma \alpha} = \Gamma C_{\alpha \beta}
\ee
or
\be
\chi_{\alpha \gamma} C_{\gamma \beta} =
\frac{\Gamma}{2} C_{\alpha \beta} .
\ee
This should hold for any conformation tensor, in particular
also for non-degenerate (invertible) tensors. This yields
\be
\chi_{\alpha \beta} = \frac{\Gamma}{2} \delta_{\alpha \beta} ,
\ee
which implies, setting an unimportant integration constant
to zero,
\be
f = \frac{\Gamma}{2} C_{\alpha \beta} \delta_{\alpha \beta}
= \frac{\Gamma}{2} C_{\alpha \alpha}
= \frac{\Gamma}{2} \textrm{tr} \tensor{C} .
\ee
However, this is not the only solution to the problem. To see this,
let us set
\be
\chi_{\alpha \beta} = \frac{\Gamma}{2} \delta_{\alpha \beta}
- \frac{\Gamma}{2} \chi^\prime_{\alpha \beta},
\ee
and insert this into Eq. \ref{eq:ConditionConservation}. This
yields
\be
0 = \frac{d {\cal H}}{dt} =
\int d \vec{r} \, \kappa_{\alpha \beta}
C_{\beta \gamma} \chi^\prime_{\gamma \alpha} .
\ee
In other words, we should find a non-trivial solution
$\tensor{\chi}^\prime$ to the matrix equation
\be
\textrm{tr} \left(
\tensor{\kappa} \cdot \tensor{C} \cdot \tensor{\chi}^\prime
\right) = 0.
\ee
Now, in the situation that $\tensor{C}$ is fully non-degenerate,
we may set
\be
\tensor{\chi}^\prime = \alpha_H (\tensor{C}) \tensor{C}^{-1} ,
\ee
with some scalar function $\alpha_H (\tensor{C})$ (the subscript
$H$ stands for ``Hamiltonian''), such that we get
\be
\alpha_H (\tensor{C}) \textrm{tr} \tensor{\kappa} = 0 ,
\ee
which does indeed hold, since $\tensor{\kappa}$ is traceless because
of incompressibility. In total, we thus find
\be
\label{eq:Result1ForChi}
\tensor{\chi} = \frac{\Gamma}{2} \left[ \tensor{1}
- \alpha_H (\tensor{C}) \tensor{C}^{-1} \right] ,
\ee
where in case of a degenerate conformation tensor we have
to set $\alpha_H = 0$.

We thus find that the conservative part of the dynamics does not
determine the Hamiltonian uniquely. Rather we find a fairly large
class of Hamiltonians which all are conserved. For a given function
$\alpha_H (\tensor{C})$, one obtains the free energy density $f$ by
integrating the relation $\partial f / \partial C_{\alpha \beta} =
\chi_{\alpha \beta}$.

\section{Hamiltonian II: Analysis of dissipative dynamics}
\label{sec:DissipativeDynamics}

The dissipative part of the equations of motion is written as
\bea
\nabla \cdot \vec{v} & = & 0 , \\
\label{eq:DissipativeStokes}
\partial_t \vec{v} & = & \nabla^2 \vec{v} , \\
\label{eq:Candidate}
\partial_t \tensor{C} & = & - \tensor{C} + \alpha_d \tensor{1} .
\eea
If we assume that the conformation tensor is non-degenerate,
we may, making use of Eq.~\ref{eq:Result1ForChi}, write the
last equation as
\bea
\partial_t \tensor{C} & = & - \frac{2}{\Gamma}
\tensor{C} \cdot \tensor{\chi} +
\left[ \alpha_d - \alpha_H (\tensor{C}) \right]
\tensor{1} ,
\\
\label{eq:UsefulFormForRelaxEq}
\partial_t C_{\alpha \beta} & = & - \frac{2}{\Gamma}
C_{\alpha \gamma} \chi_{\gamma \beta} +
\left[ \alpha_d - \alpha_H (\tensor{C}) \right]
\delta_{\alpha \beta} .
\eea
We now multiply Eq.~\ref{eq:DissipativeStokes} with $\vec{v}$,
integrate over space, and apply integration by parts, to obtain the
standard viscous dissipation rate
\be
\label{eq:ViscousDissipationRate}
\frac{d}{dt} \int d \vec{r} \, \frac{\vec{v}^2}{2} =
- \int d \vec{r} \, \left( \partial_\beta v_\alpha \right)
\left( \partial_\beta v_\alpha \right) \le 0 .
\ee
Similarly, multiplying Eq.~\ref{eq:UsefulFormForRelaxEq} with
$\chi_{\alpha \beta}$ yields
\be
\label{eq:DissipationRateFromChi}
\partial_t f = - \frac{2}{\Gamma} \chi_{\beta \alpha}
C_{\alpha \gamma} \chi_{\gamma \beta} +
\left[ \alpha_d - \alpha_H (\tensor{C}) \right]
\chi_{\alpha \alpha} .
\ee
Now, the Second Law of thermodynamics means that the
dynamics should satisfy
\be
\frac{d {\cal H}}{dt} \le 0 .
\ee
From Eqs.~\ref{eq:ViscousDissipationRate} and
\ref{eq:DissipationRateFromChi} one sees that the only term that
causes a potential violation of the Second Law is the second term of
Eq.~\ref{eq:DissipationRateFromChi} --- note that the first term is
strictly non-positive, due to the positive-semidefiniteness of
$\tensor{C}$. In other words, the condition for the Second Law is
\be
\label{eq:ConditionTraceChi}
\left[ \alpha_d - \alpha_H (\tensor{C}) \right]
\textrm{tr} \tensor{\chi} \le 0 .
\ee
One possibility to achieve that goal is to set $\alpha_H(\tensor{C})
= \alpha_d$, in which case the term simply vanishes. In other words,
we may thus pick, out of the large class of possible Hamiltonians,
one particular Hamiltonian, for which indeed the Second Law holds.
We then obtain
\bea
\tensor{\chi} & = & 
\frac{\Gamma}{2} \left[ \tensor{1}
- \alpha_d \tensor{C}^{-1} \right] ,
\\
f & = & \frac{\Gamma}{2} \left[ \textrm{tr} \tensor{C}
  - \alpha_d \textrm{tr} \ln \tensor{C} \right] ,
\eea
where we have set an unimportant integration constant to zero.  For
$\alpha_d > 0$, this Hamiltonian is bounded from below, as it should
be. For $\alpha_d = 0$, this is also the case, due to the
positive-semidefiniteness of $\tensor{C}$.

We can thus always satisfy the Second Law by choosing an appropriate
Hamiltonian. If one then has additional arguments, e.~g. from a
microscopic model, why that particular Hamiltonian should be the
physically correct one, then the latter can be used as well. Indeed,
this is the route of the standard Oldroyd-B model: Here one derives
the Hamiltonian with $\alpha_d = 1$ from a microscopic model, and the
equation of motion (again with $\alpha_d = 1$) from the corresponding
kinetic theory. The main point of this paper is however that this
choice is less compelling than one might think. The choice $\alpha_d =
0$ is, from the microscopic point of view, physically acceptable as
well, and, in our opinion, probably even preferable. We will outline
these arguments in more detail below.

Are there other solutions of the problem posed in
Eq.~\ref{eq:ConditionTraceChi}? For arbitary functions $\alpha_H
(\tensor{C})$ this analysis is difficult, but for the case that
$\alpha_H$ is simply a constant independent of $\tensor{C}$ it is
fairly easy. If $\lambda_k > 0$ are the eigenvalues of $\tensor{C}$,
then
\be
\left( \alpha_d - \alpha_H \right)
\textrm{tr} \tensor{\chi}
=
\frac{\Gamma}{2}
\left( \alpha_d - \alpha_H \right)
\sum_k \left( 1 - \alpha_H \lambda_k^{-1} \right) .
\ee
Assuming $\alpha_d \ne \alpha_H$, and also $\alpha_H > 0$ (negative
values are prohibited, since then the Hamiltonian were not bounded
from below), we see that this expression does \emph{not} have a
definitive sign, since $1 - \alpha_H \lambda_k^{-1}$ changes sign upon
varying $\lambda_k$. The only exception is $\alpha_H = 0$, in which
case the expression does not depend on the conformation tensor at
all. In this case, however, $\alpha_d \le 0$ is needed for the Second
Law. Since on the other hand we need $\alpha_d \ge 0$ for maintaining
the positivity of $\tensor{C}$, the only possibility that remains is
$\alpha_d = 0$, which brings us back to the original case $\alpha_d =
\alpha_H$.

To summarize our results so far: We have discussed a class of
generalized Oldroyd-B models, with constitutive equation
\be
\delta_t \tensor{C} = - \tensor{C} + \alpha_d \tensor{1} ,
\ee
where $\alpha_d \ge 0$, such that the dynamics conserves the positive
semi-definiteness of $\tensor{C}$. We have seen that the conservative
part of the dynamics conserves the Hamiltonian
\be
{\cal H} = \int d \vec{r} \,
\left[ \frac{\vec{v}^2}{2} + f (\tensor{C}) \right] ,
\ee
with
\be
\label{eq:FreeEnergyDensityMacroscopic}
f (\tensor{C}) =  \frac{\Gamma}{2} \left[ \textrm{tr} \tensor{C}
  - \alpha_H \textrm{tr} \ln \tensor{C} \right] ,
\ee
where $\alpha_H \ge 0$ to make sure that the Hamiltonian is bounded
from below. Furthermore, we have seen that the Second Law requires
that $\alpha_d = \alpha_H = \alpha$ (we will omit the indexes $d$ and
$H$ from now on). In this case the constitutive equation can be
written in the canonical form $\delta_t \tensor{C} = $ ``transport
coefficient times thermodynamic driving force'', where the negative of
the latter is given by
\be
\tensor{\chi} = \frac{\partial f}{\partial \tensor{C}} =
\frac{\Gamma}{2} \left[ \tensor{1}
- \alpha \tensor{C}^{-1} \right] ,
\ee
and the constitutive equation reads
\be
\delta_t \tensor{C} = - \frac{2}{\Gamma}
\tensor{C} \cdot \tensor{\chi} .
\ee
Similarly, the dissipation rate assumes the canonical expression
required by the GENERIC
formalism~\cite{grmelaDynamicsThermodynamicsComplex1997,
  ottingerDynamicsThermodynamicsComplex1997,
  ottingerEquilibriumThermodynamics2005}, i.~e. a quadratic form in
the driving forces:
\be
\label{eq:QuadraticFormDissipation}
\partial_t f = - \frac{2}{\Gamma} \textrm{tr}
\left( \tensor{\chi} \cdot \tensor{C} \cdot
\tensor{\chi} \right) .
\ee
The basic principles of equilibrium and non-equilibrium thermodynamics
are thus satisfied, and this is true for any value $\alpha \ge 0$. The
standard Oldroyd-B model is the special case $\alpha = 1$.

Interestingly, for $\alpha > 0$ the minimal Hamiltonian and also the
minimum dissipation rate do not occur at $\tensor{C} = 0$ but rather
at $\tensor{C} = \alpha \tensor{1}$, where $\tensor{\chi} = 0$.

An even more interesting observation is that, for $\alpha > 0$, the
dissipation rate becomes very large whenever the eigenvalues of
$\tensor{C}$ become either large or small. Actually, the rate tends to
infinity when one of the eigenvalues tends to zero. This may perhaps
provide an explanation for the numerical evidence obtained in
Ref.~\onlinecite{bajajCoilstretchTransitionBreakdown2008}, which
considered the flow of a (standard, $\alpha = 1$) Oldroyd-B fluid
around an obstacle. The data seem to indicate that a solution beyond a
certain critical Weissenberg number $Wi$ simply does not exist (``high
Weissenberg number problem'', HWNP). In this context, it should be
recalled that $Wi$ is defined as the dimensionless product of shear
rate and molecular relaxation time. In other words, no solution seems
to exist at high flow rates. Now, from a physical point of view, we
must assume that there is some ``pump'' that provides a constant power
to move the fluid and to balance the dissipative losses, such that a
net non-equilibrium steady state is established. If the system
provides an infinite energy sink, it is conceivable that we run into a
situation where we keep on increasing the pumping power without the
flow rate increasing any further, such that higher values of $Wi$
become inaccessible.

\emph{This situation is significantly different if we set $\alpha =
0$.} Here we have $f = (\Gamma / 2) \textrm{tr} \tensor{C}$ and
$\partial_t f = - (\Gamma / 2) \textrm{tr} \tensor{C} = -f$. We still
get large energies and large dissipation rates for large eigenvalues,
but the singularity at $\tensor{C} = 0$ is removed. We therefore
propose in the present paper to consider the new constitutive
equation, which results from setting $\alpha = 0$. It may be that this
model has perhaps a less severe, or perhaps even non-existing,
HWNP. Numerical investigation of the behavior of this system is left
for future work. In the present paper, we wish to establish that this
model is thermodynamically consistent (this is the content of the
paper so far), and that it is \emph{not} an arbitrary \emph{ad hoc}
choice, but can be derived from the same microscopic picture upon
which the standard Oldroyd-B model is based. The key to construct this
new constitutive equation is \emph{to do the statistical mechanics in
a different ensemble}.

\section{General considerations on micro-macro coupling}
\label{sec:MicroMacro}

\subsection{General remarks}
\label{sec:GeneralRemarks}

The basic idea behind the standard derivation of the Oldroyd-B model
from kinetic theory may be sketched as follows. Starting points are
the following observations and assumptions: (i) The influence of the
polymer degrees of freedom on the flow behavior enters only via the
term $\Gamma \nabla \cdot \tensor{C}$ of the momentum equation.
Therefore one needs an equation of motion for $\tensor{C}$. (ii)
Physically, $\tensor{C}$ is nothing but the (properly normalized)
stress coming from the polymers. (iii) On a microscopic level, the
polymer chains should be represented by some sort of bead-spring
model. (iv) Within such a model, the Kramers (or virial) formula
provides a well-defined expression for the stress in terms of bead
coordinates. (v) This stress tensor must then be averaged, and the
average value enters the macroscopic momentum eqution. (vi) At the
same time, the underlying microscopic model for the dynamics results
in an equation of motion for the averaged stress, which is then the
desired constitutive equation.

The Oldroyd-B model then postulates very specific assumptions on (i)
the microscopic model, (ii) its dynamics, and (iii) \emph{the
averaging procedure}. The combination of these three ingredients then
gives rise to the specific formulation of the model.

The point that we wish to make here is that the averaging procedure is
much less unique and compelling than one might think. As a matter of
fact, there are several choices possible, and we wish to present here
a new alternative way for the averaging procedure, which then gives
rise to the modified constitutive equation with $\alpha = 0$.
Conversely, the microscopic model, both in terms of its degrees of
freedom, and in terms of its assumed dynamics, will remain untouched.

\subsection{Constrained averages}
\label{sec:Constrained}

The lack of uniqueness becomes most transparent if formulated in a
general and abstract language. We assume that the microscopic
description is based on a vector of microscopic phase-space
coordinates $\vec{\xi}$, and that the dynamics can be described in
terms of the evolution of the phase-space probability density
$P(\vec{\xi},t)$. For an observable $A$ (i.~e. a phase-space function
$A(\vec{\xi})$), the simple thermal average is then given by
\be
\left< A \right> (t) = \int d \vec{\xi}
P(\vec{\xi},t) A(\vec{\xi}) .
\ee
However, this average is typically \emph{not} the average whose result
should be fed into the macroscopic description. As is well-known from
the Mori-Zwanzig
formalism~\cite{forsterHydrodynamicFluctuationsBroken1995,
  j.phansenTheorySimpleLiquids2013}, the macroscopic description is
rather based upon the identification (or better: choice) of a set of
macroscopic or ``slow'' observables $X_i$, i.~e. functions
$X_i(\vec{\xi})$, which facilitate the coupling. Averaging should
then only be done over the remaining ``fast'' variables.

In order to exclude the slow variables from averaging, we introduce
Dirac delta functions in the phase space integrals, analogously to the
microcanonical ensemble known from standard statistical physics. We
therefore define the constrained average of an observable
$A(\vec{\xi})$ as follows:
\bea
\nonumber
\left[ A \right] (t) & = &
\frac{
\int d \vec{\xi} \, \prod_j \delta (X_j - Y_j) P A}
{\int d \vec{\xi} \, \prod_j \delta (X_j - Y_j) P}
\\
& = &
\frac{\left< \prod_j \delta (X_j - Y_j) A \right>}
{\left< \prod_j \delta (X_j - Y_j) \right>} ,
\eea
where $Y_j$ is the macroscopic value of $X_j$. To determine the $Y_j$,
we postulate the reasonable requirement that they should be the
unconstrained averages of the $X_j$:
\be
Y_j = \left< X_j \right> .
\ee
Furthermore, we have
\bea
\nonumber
\left[ X_i \right]
& = &
\frac{\left< \prod_j \delta (X_j - Y_j) X_i \right>}
{\left< \prod_j \delta (X_j - Y_j) \right>}
\\
& = &
\frac{\left< \prod_j \delta (X_j - Y_j) Y_i \right>}
{\left< \prod_j \delta (X_j - Y_j) \right>}
= Y_i ,
\eea
as it should be.

One thus sees that one gets different averaging procedures depending
on the choice of the slow variables $X_i$. Traditional polymer
rheology has always assumed that the only reasonable choice for the
$X_i$ are the components of the stress tensor, or, directly related to
it, the components of the conformation tensor. However, the
conformation tensor may as well take the role of an observable $A$,
while for the $X_i$ different variables are being used. What we
propose in the present paper is to rather take the components of the
end-to-end vector for the $X_i$.

As a shorthand notation, we combine the set of observables $X_i$ and
their averages $Y_i$ as vectors $\vec{X}$ and $\vec{Y}$. Similarly, we
assume that we are not interested in only a single observable $A$, but
in a whole set, again written as a vector $\vec{A}$. We thus may write
\be
\label{eq:ConstrainedAverageVectorForm}
\left[ \vec{A} \right] = \frac{
  \left< \delta \left( \vec{X} - \vec{Y} \right) \vec{A} \right>
  }{
  \left< \delta \left( \vec{X} - \vec{Y} \right) \right>
  }
\ee
and
\be
\left[ \vec{X} \right] = \vec{Y} = \left< \vec{X} \right> . 
\ee

\subsection{Hamiltonian}
\label{sec:HamiltonianGeneralStatMech}

One should note that the macroscopic Hamiltonian should \emph{not} be
calculated according to that recipe --- it is not an observable
but rather a thermodynamic potential. Assuming that the microscopic
system is governed by a Hamiltonian ${\cal H}_0 (\vec{\xi})$, and
is studied in the canonical ensemble, where $\beta = 1 / (k_B T)$
($k_B$ Boltzmann's constant, $T$ absolute temperature), then the
constrained partition function is
\be
Z \left( \vec{Y} \right) =
\int d \vec{\xi} \delta \left( \vec{X} - \vec{Y} \right)
\exp \left( - \beta {\cal H}_0 \right) ,
\ee
and the macroscopic Hamiltonian (or free energy) is
\be
{\cal H} = - \beta^{-1} \ln Z.
\ee

\subsection{Fokker-Planck dynamics}
\label{sec:FPEGeneral}

We now assume that the underlying dynamics on the microscale
is described by an evolution equation of the Fokker-Planck
type for the probability density $P$,
\be
\partial_t P = {\cal L} P ,
\ee
where ${\cal L}$ is the Fokker-Planck operator
\be
{\cal L} =
\frac{\partial}{\partial \vec{\xi}}
\cdot \tensor{\cal D} \cdot
\frac{\partial}{\partial \vec{\xi}} -
\frac{\partial}{\partial \vec{\xi}}
\cdot \vec{V} ;
\ee
here $\tensor{\cal D}$ denotes the (symmetric and
positive-semidefinite) diffusion tensor, and $\vec{V}$ the drift
velocity in $\vec{\xi}$ space. Its adjoint is written as
\be
{\cal L}^\dag =
\frac{\partial}{\partial \vec{\xi}}
\cdot \tensor{\cal D} \cdot
\frac{\partial}{\partial \vec{\xi}} +
\vec{V} \cdot
\frac{\partial}{\partial \vec{\xi}} .
\ee
For the time evolution of the unconstrained average of an
observable $A$ we thus find
\bea
\nonumber
\partial_t \left< A \right>
& = &
\partial_t \int d\vec{\xi} \, A(\vec{\xi}) P(\vec{\xi}, t)
\\
\nonumber
& = &
\int d\vec{\xi} \, A(\vec{\xi}) {\cal L}
P(\vec{\xi}, t)
\\
\nonumber
& = &
\int d\vec{\xi} \, P(\vec{\xi}, t)
{\cal L}^\dag A(\vec{\xi})
\\
& = &
\left< {\cal L}^\dag A \right> .
\eea
This, in turn, tells us that the time evolution of the vector
of ensemble-defining quantities $\vec{Y}$ is given by
\be
\label{eq:GeneralConstitutiveEq}
\partial_t \vec{Y} = \partial_t \left< \vec{X} \right>
= \left< {\cal L}^\dag \vec{X} \right> .
\ee
We thus find that the rheological model is constructed from the
combination of the microscopic Fokker-Planck model with a set of
chosen ensemble-defining observables $\vec{X}$. Equation
\ref{eq:GeneralConstitutiveEq} (or a suitable equivalent formulation)
thus turns out to be the \emph{constitutive equation} of the
rheological model.

One should realize that one obtains a closed-form constitutive
equation only if ${\cal L}^\dag \vec{X}$ is expressable as a linear
function of $\vec{X}$. For the Oldroyd-B models (both standard and
modified) discussed in the present paper this is the case; however in
general this is more the exception than the rule. If it is not the
case, Eq.~\ref{eq:GeneralConstitutiveEq} nevertheless remains not only
valid (within the framework of the chosen model), but also practically
useful: It is always possible to estimate the RHS of
Eq.~\ref{eq:GeneralConstitutiveEq} by stochastic Brownian Dynamics
(BD) simulations. This philosophy has already been put into practice
in the so-called CONNFFESSIT
approach~\cite{lasoCalculationViscoelasticFlow1993}.

The micro-macro coupling, however, involves not only the vector of
ensemble-defining quantities $\vec{X}$, but also a vector of
observables $\vec{A}$, whose constrained average $\left[ \vec{A}
  \right]$ appears in the macroscopic dynamic equation. If $\vec{A}$
happens to coincide with $\vec{X}$, then matters are easy, since then
\be
\left[ \vec{A} \right] = \left[ \vec{X} \right] = \vec{Y} ,
\ee
such that the results from the integration of the constitutive
equation may be used directly. This is exactly the route that is taken
in the standard variant of the Oldroyd-B model. However, in the
general case one rather has to use
Eq.~\ref{eq:ConstrainedAverageVectorForm}. The difficulty here is that
in many cases it is not possible to construct a closed-form expression
for the RHS of Eq.~\ref{eq:ConstrainedAverageVectorForm}. This is
directly related to the fact that one typically cannot construct a
closed-form analytical solution of the Fokker-Planck equation (FPE),
if one allows for arbitrary initial conditions and arbitrary
non-equilibrium external driving with unknown time dependence. We
presume, however, that it should be possible to construct a suitable
sampling algorithm which allows the estimation of the averages on the
RHS of Eq.~\ref{eq:ConstrainedAverageVectorForm} by BD. The details of
such a procedure are however not yet clear and are a topic for future
research.

There is, however, yet one other case where the evaluation of $\left[
  \vec{A} \right]$ becomes very easy. This is if the constraints apply
to the whole underlying set of dynamic variables, such that we simply
have $\vec{X} = \vec{\xi}$ and no non-trivial averaging
remains. Indeed, we then have
\bea
\nonumber
\left< \delta \left( \vec{X} - \vec{Y} \right) \vec{A} \right>
& = &
\int d\vec{\xi} \delta \left( \vec{\xi} - \vec{Y} \right)
\vec{A} \left( \vec{\xi} \right) P \left( \vec{\xi}, t \right)
\\
& = &
\vec{A} \left( \vec{Y} \right) P \left( \vec{Y}, t \right)
\eea
and
\be
\left< \delta \left( \vec{X} - \vec{Y} \right) \right>
=
P \left( \vec{Y}, t \right) ,
\ee
such that
\be
\left[ \vec{A} \right] = \vec{A} \left( \vec{Y} \right) ,
\ee
which allows us to again directly use the results of the integration
of the constitutive equation. It is this second route that we use in
the present paper to construct the modified Oldroyd-B model.

\section{The Oldroyd-B Hamiltonian from statistical mechanics}
\label{sec:StandardHamilStatMech}

From now on, we will use conventional units again. Consider a single
Hookean dumbbell (this is the simplified model for a polymer molecule)
with spring constant $k$ in $d$-dimensional space. In thermal
equilibrium, the vector $\vec{q}$, which connects the two beads with
each other, will have a mean square extension (in tensorial form) of
\be
\label{eq:ExtVect2ndMomentThermalEq}
\left< q_\alpha q_\beta \right>
= \frac{k_B T}{k} \delta_{\alpha \beta} ,
\ee
as a direct consequence of the equipartition theorem. This motivates
the introduction of a non-dimensionalized extension vector $\vec{p}$
via
\be
\vec{q} = \left( \frac{k_B T}{k} \right)^{1/2} \vec{p} ,
\ee
such that
\be
\left< p_\alpha p_\beta \right> = \delta_{\alpha \beta} .
\ee
Consider now a set of $N$ such dumbbells with normalized extension
vectors $\vec{p}_i$. Assuming that there is no interaction whatsoever,
except the intramolecular spring forces, the corresponding Hamiltonian
is given by
\be
\beta {\cal H}_0 = \frac{1}{2} \sum_i \vec{p}_i^2 .
\ee

\subsection{Conformation tensor ensemble}
\label{sec:HamiltonianConformationTensorEnsemble}

Let us now define a microscopic expression for the conformation
tensor via
\be
\hat{C}_{\alpha \beta} = N^{-1} \sum_i p_{i \alpha} p_{i \beta} .
\ee
Consequently,
\be
\beta {\cal H}_0 = \frac{N}{2} \hat{C}_{\alpha \alpha}
= \frac{N}{2} \textrm{tr} \hat{\tensor{C}} .
\ee
We now wish to evaluate the partition function at fixed conformation
tensor $\tensor{C}$, of which we assume that it is non-degenerate. A
fixed conformation tensor means a fixed value for ${\cal H}_0$,
i.~e. $\beta {\cal H}_0 = (N/2) \textrm{tr} \tensor{C}$. The condition
that the configuration of dumbbells, $\{ \vec{p}_i \}$, has to satisfy
is therefore
\be
N^{-1} \sum_i \vec{p}_i \vec{p}_i^T = \tensor{C} .
\ee
Introducing the variable transformation
\be
\vec{p}_i = \tensor{C}^{1/2} \vec{\pi}_i ,
\ee
the condition may also be written as
\be
\tensor{C}^{1/2}
N^{-1} \sum_i \vec{\pi}_i \vec{\pi}_i^T
\tensor{C}^{1/2} = \tensor{C} ,
\ee
or, after multiplication with $\tensor{C}^{-1/2}$ from both
left and right,
\be
N^{-1} \sum_i \vec{\pi}_i \vec{\pi}_i^T = \tensor{1} .
\ee
Furthermore, we notice
\be
d \vec{p}_1 d \vec{p}_2 \ldots d \vec{p}_N =
\left[ \det \tensor{C}^{1/2} \right]^N
d \vec{\pi}_1 d \vec{\pi}_2 \ldots d \vec{\pi}_N
\ee
and re-write
\bea
\nonumber
\left[ \det \tensor{C}^{1/2} \right]^N
& = &
\left[ \det \tensor{C} \right]^{N/2}
=
\exp \left( \frac{N}{2} \ln \det \tensor{C} \right)
\\
& = &
\exp \left( \frac{N}{2} \textrm{tr} \ln \tensor{C} \right) .
\eea
Similarly, we find
\bea
\nonumber
&&
\delta \left( N^{-1} \sum_i \vec{p}_i \vec{p}_i^T - \tensor{C} \right)
\\
\nonumber
& = &
\delta \left( \tensor{C}^{1/2} \left(
N^{-1} \sum_i \vec{\pi}_i \vec{\pi}_i^T - \tensor{1}
\right) \tensor{C}^{1/2} \right)
\\
\nonumber
& = &
\left( \det \tensor{C} \right)^{-1} \delta \left(
N^{-1} \sum_i \vec{\pi}_i \vec{\pi}_i^T - \tensor{1}
\right)
\\
& = &
\exp \left( - \textrm{tr} \ln \tensor{C} \right)
\delta \left(
N^{-1} \sum_i \vec{\pi}_i \vec{\pi}_i^T - \tensor{1}
\right) .
\eea
All in all, this results in
\be
Z = \exp \left( - \frac{N}{2} \textrm{tr} \tensor{C} +
\frac{N - 2}{2} \textrm{tr} \ln \tensor{C} \right) \tilde{Z} ,
\ee
with
\be
\tilde{Z} =
\int d \vec{\pi}_1 d \vec{\pi}_2 \ldots d \vec{\pi}_N
\delta \left( 
N^{-1} \sum_i \vec{\pi}_i \vec{\pi}_i^T - \tensor{1}
\right) .
\ee
$\tilde{Z}$ does not depend on $\tensor{C}$ anymore, which means that
it may be set to unity, by means of choosing a proper normalization of
the partition function (or, equivalently, a convenient zero for the
free energy). We thus find
\bea
\nonumber    
{\cal H}
& = &
k_B T \frac{N}{2} \textrm{tr} \tensor{C}
- k_B T \frac{N - 2}{2} \textrm{tr} \ln \tensor{C}
\\
& \approx &
k_B T \frac{N}{2} \left(
\textrm{tr} \tensor{C} - \textrm{tr} \ln \tensor{C}
\right) ,
\eea
where in the last step we have assumed that $N$ is large. Assuming
that the system is confined to a volume $V$, such that $n = N/V$ is
the number of dumbbells per unit volume, the free energy per unit
volume is
\be
f \left( \tensor{C} \right) =
k_B T \frac{n}{2} \left( \textrm{tr} \tensor{C}
- \textrm{tr} \ln \tensor{C} \right) .
\ee
This is indeed the free energy of the standard Oldroyd-B model as
identified before, see Eq. \ref{eq:FreeEnergyDensityMacroscopic} with
$\alpha = 1$, where we need to identify $n k_B T$, after
transformation to natural units, with the parameter $\Gamma$.

\subsection{End-to-end vector ensemble}
\label{sec:HamiltonianEndToEndVectorEnsemble}

We now wish to perform the same exercise as in the previous
subsection; however this time we do not wish to keep the conformation
tensor fixed but rather the normalized end-to-end vector
\be
\hat{\vec{Q}} = N^{-1} \sum_i \vec{p}_i .
\ee
This problem can be solved very easily, relying on standard results of
Gaussian statistics. We write the partition function with a
constraining value $\vec{Q}$ as
\be
Z = \frac{
  \int d\vec{p}_1 \ldots d\vec{p}_N
  \delta \left( \vec{Q} - \hat{\vec{Q}} \right)
  \exp \left( - \beta {\cal H}_0 \right)
}{
  \int d\vec{p}_1 \ldots d\vec{p}_N
  \exp \left( - \beta {\cal H}_0 \right)
} ,
\ee
where we have introduced the denominator for convenient
normalization. Because of $\int d \vec{Q} Z(\vec{Q}) = 1$ we may
interpret $Z$ just as the probability density for the end-to-end
vector. However, the $\vec{p}_i$ are just Gaussian random variables
and $\hat{\vec{Q}}$ is a linear combination thereof, meaning that it
is Gaussian as well. Trivial evaluation yields
\be
\left< \vec{Q} \right> = 0
\ee
and
\be
\left< Q_\alpha Q_\beta \right> = N^{-1} \delta_{\alpha \beta} ,
\ee
which means
\be
Z (\vec{Q}) = \textrm{const.}
\exp \left( - \frac{N}{2} \vec{Q}^2 \right) .
\ee
After re-adjusting the zero of the free energy, we thus obtain
\be
{\cal H} = k_B T \frac{N}{2} \vec{Q}^2 .
\ee
Within the framework of this modified theory, we therefore
have to define the conformation tensor differently ---
instead of $\hat{\tensor{C}} = N^{-1} \sum_i \vec{p}_i \vec{p}_i^T$
(``average of the square''), we now have to consider
\be
\hat{\tensor{C}}_Q =
\hat{\vec{Q}} \hat{\vec{Q}}^T =
\left( N^{-1} \sum_i \vec{p}_i \right)
\left( N^{-1} \sum_i \vec{p}_i \right)^T
\ee
(``square of the average''). With that re-definition of the
conformation tensor, we get
\be
{\cal H} = k_B T \frac{N}{2} \textrm{tr} \tensor{C}_Q
\ee
and for the free energy per unit volume
\be
f \left( \tensor{C}_Q \right) =
k_B T \frac{n}{2} \textrm{tr} \tensor{C}_Q ,
\ee
which is identical to the result derived before
(Eq. \ref{eq:FreeEnergyDensityMacroscopic}), however this time with
$\alpha = 0$.

\section{Kramers stress tensor}
\label{sec:Kramers}

The key to couple the Fokker-Planck system to the hydrodynamics of the
solution is the Kramers (or virial) expression for the stress
tensor~\cite{j.phansenTheorySimpleLiquids2013}. Again we consider our
set of $N$ harmonic dumbbells immersed homogeneously in a volume
$V$. The microscopic expression for the polymer part of the stress
tensor is then
\be
\tau^{(mic)}_{\alpha \beta} =
- \frac{N}{V} k_B T \delta_{\alpha \beta}
- \frac{1}{V} \sum_{i} q_{i \alpha} F_{i \beta} .
\ee
Here $\vec{q}_{i}$ is the vector that connects the two beads that
together form the dumbbell $i$ under consideration. Similarly,
$\vec{F}_{i}$ is the corresponding force,
$\vec{F}_{i} = - k \vec{q}_{i}$. Therefore,
\be
\tau^{(mic)}_{\alpha \beta} =
- \frac{N}{V} k_B T \delta_{\alpha \beta}
+ \frac{k}{V} \sum_{i} q_{i \alpha} q_{i \beta} .
\ee
Again we introduce normalized connector vectors $\vec{p}_{i}$ via
\be
\vec{q}_{i} =
\left( \frac{k_B T}{k} \right)^{1/2} \vec{p}_{i} ,
\ee
which gives rise to
\bea
\nonumber
\tau^{(mic)}_{\alpha \beta}
& = &
- \frac{N}{V} k_B T \delta_{\alpha \beta}
+ \frac{k_B T}{V} \sum_{i} p_{i \alpha} p_{i \beta}
\\
\nonumber
& = &
- \frac{N}{V} k_B T \delta_{\alpha \beta}
+ \frac{N k_B T}{V} \hat{C}_{\alpha \beta}
\\
& = &
n k_B T \left( \hat{C}_{\alpha \beta} -
\delta_{\alpha \beta} \right) .
\eea

On the macroscopic (hydrodynamic) level, an analogous macroscopic
version of $\tensor{\tau}$ will enter. It is clear that this must be
some suitably averaged stress tensor. As we have learned from the
previous developments, the averaging depends on the underlying
ensemble, or, in other words, on the choice of the ``slow'' variables.
We thus obtain
\be
\tau_{\alpha \beta} =
n k_B T \left( \left[ \hat{C}_{\alpha \beta} \right] -
\delta_{\alpha \beta} \right) .
\ee
As the second term does not contribute to the forcing term in the
hydrodynamic equation of motion, it may as well be omitted. This
yields
\be
\tau_{\alpha \beta} =
n k_B T \left[ \hat{C}_{\alpha \beta} \right] .
\ee
We have already learned that in dimensionless units we need to
identify $n k_B T$ with $\Gamma$. In summary, we thus obtain a
rheological model which is specified by the momentum equation
(in reduced units)
\bea
\nabla \cdot \vec{v} & = & 0 , \\
D_t \vec{v} & = & - \nabla p + \nabla^2 \vec{v}
+ \Gamma \nabla \cdot \left[ \hat{\tensor{C}} \right] ,
\eea
augmented by a constitutive equation, i.~e. an equation of motion for
$\left[ \hat{\tensor{C}} \right]$. To make further progress, we need
to consider the microscopic dynamics in detail (see the following
section), and also the details of the constrained average. As all
dumbbells have the same properties, we may write both the constrained
and the unconstrained averages of $\hat{\tensor{C}}$ as
single-dumbbell averages:
\bea
\left[ \hat{C}_{\alpha \beta} \right] & = &
\left[ p_\alpha p_\beta \right] , \\
\left< \hat{C}_{\alpha \beta} \right> & = &
\left< p_\alpha p_\beta \right> .
\eea

For the ensemble which fixes the conformation tensor (standard
Oldroyd-B model), we can directly take advantage of the fact that for
ensemble-defining observables the constrained average coincides with
the unconstrained one, see Sec.~\ref{sec:MicroMacro}, such that
$\left[ \hat{\tensor{C}} \right] = \left< \hat{\tensor{C}}
\right>$. In other words, for the standard Oldroyd-B model we
have
\be
\left[ \hat{C}_{\alpha \beta} \right] = 
\left[ p_\alpha p_\beta \right] =
\left< \hat{C}_{\alpha \beta} \right> =
\left< p_\alpha p_\beta \right> .
\ee
This means that we just have to find the equation of motion for
$\left< p_\alpha p_\beta \right>$ within a single-dumbbell picture,
and the construction of the model is done. This is an easy and
straightforward task, and this is the theoretical reason why the
standard version of the Oldroyd-B model is so popular.

If instead the average end-to-end vector $\vec{Q}$ defines the
ensemble, things are slightly more involved. We here have the
situation where the observables that are needed for the macroscopic
equations differ from those that define the ensemble. As already
outlined at the end of Sec.~\ref{sec:MicroMacro}, it is in that
situation typically impossible to construct a closed constitutive
equation of motion, except if the number of constraints is identical
to the number of degrees of freedom. We are therefore led to the
conclusion that in the present case we should set $N = 1$, such
that the constraint acts on the single-dumbbell end-to-end vector.
In this case, we trivially obtain
\be
\left[ p_\alpha p_\beta \right] = Q_\alpha Q_\beta .
\ee
Since on the other hand $Q_\alpha = \left< p_\alpha \right>$, we
may also write
\be
\left[ p_\alpha p_\beta \right] =
\left< p_\alpha \right> \left< p_\beta \right> .
\ee
Again, we are led to consider the square of the average instead of the
average of the square (a similar notion already occurred when we
calculated the Hamiltonian in
Sec.~\ref{sec:HamiltonianEndToEndVectorEnsemble}). The desired
constitutive equation therefore is found from a single-dumbbell
dynamical model, by finding the equation of motion for $\left<
p_\alpha \right> \left< p_\beta \right>$, which again is an easy task.

The fact that one should set $N = 1$ is further corroborated
by a consideration of the averages \emph{in thermal equilibrium},
which can be calculated exactly, for arbitrary values of $N$,
by straightforward evaluation of Gaussian integrals. One
finds (see Appendix \ref{sec:GaussianAverage})
\be
\label{eq:ConstrainedAverageConfTensorGaussian}
\left[ p_{\alpha} p_{\beta} \right]
=
Q_\alpha Q_\beta + \left( 1 - N^{-1} \right) \delta_{\alpha \beta} ,
\ee
which may be written in the suggestive form
\bea
\nonumber
&&
\left[ p_{\alpha} p_{\beta} \right]
-
\left[ p_{\alpha} \right] \left[p_{\beta} \right]
\\
& = &
\left( 1 - N^{-1} \right)
\left( \left< p_{\alpha} p_{\beta} \right>
-
\left< p_{\alpha} \right> \left<p_{\beta} \right> \right) .
\eea
In other words, the constraint reduces the fluctuations by the factor
$1 - N^{-1}$. For $N = 1$ there are no fluctuations left (this result
was already clear from the considerations above), while for $N \to
\infty$ there is no reduction in fluctuations whatsoever. This is in
perfect accordance with the general notion in statistical physics that
in the thermodynamic limit ($N \to \infty$) constraints do not matter,
and ensembles become equivalent. It is reasonable to assume (though
difficult to prove) that similar behavior also occurs in
nonequilibrium situations. We thus see that, in order to bring about a
sizable effect of the constraint, one has to consider a small value of
$N$, meaning in practice $N = 1$. We feel that such a constrained
ensemble acting on the single-dumbbell distribution is not physically
unreasonable.

\section{Fokker-Planck equation}
\label{sec:FPE}

As already discussed in the previous section, for our purposes a
single-dumbbell picture is sufficient. We thus consider a single
Hookean dumbbell with connector vector $\vec{q}$ and normalized
connector vector $\vec{p}$. If $\vec{r}$ denotes the center of mass of
the dumbbell, then the beads are located at the positions $\vec{r} \pm
\vec{q} / 2$, and the velocity flow field at the two bead positions is
$\vec{v} (\vec{r}) \pm \tensor{\kappa} (\vec{r}) \cdot \vec{q} /
2$. Here we neglect higher-than-linear terms in the velocity profile,
which is reasonable, given the smallness of polymer
molecules. Assuming stick boundary conditions for the beads,
i.~e. assuming that the beads just move with the flow, we find
$\dot{\vec{q}} = \tensor{\kappa} (\vec{r}) \cdot \vec{q}$. Adding the
effect of the spring force, and thermal noise, the corresponding
Langevin equation for $\vec{q}$ is
\be
\dot{q}_\alpha = \kappa_{\alpha \beta} q_\beta
- \mu k q_\alpha + \eta_\alpha ,
\ee
where $\mu$ is a mobility, and $\eta_\alpha$ is a Gaussian white
noise with
\bea
\left< \eta_\alpha (t) \right> & = & 0 , \\
\left< \eta_\alpha (t) \eta_\beta (t') \right> & = &
2 \mu k_B T \delta_{\alpha \beta} \delta (t - t') .
\eea
In terms of $\vec{p}$, this is written as
\be
\dot{p}_\alpha = \kappa_{\alpha \beta} p_\beta
- \frac{1}{\tau} p_\alpha + \xi_\alpha ,
\ee
where the relaxation time is given by $\tau = (\mu k)^{-1}$ and
\bea
\left< \xi_\alpha (t) \right> & = & 0 , \\
\left< \xi_\alpha (t) \xi_\beta (t') \right> & = &
2 \frac{1}{\tau} \delta_{\alpha \beta} \delta (t - t') .
\eea
If $P(\vec{r}, \vec{p}, t)$ denotes the probability density
in $\vec{p}$ space, with
\be
\int d \vec{p} \, P(\vec{r}, \vec{p}, t) = 1 ,
\ee
then $P$ obeys the Fokker-Planck equation (derived from the
above Langevin equation)
\be
\partial_t P = {\cal L} P ,
\ee
where ${\cal L}$ is the Fokker-Planck operator, which has
the explicit form
\bea
\nonumber
{\cal L} & = & - \frac{\partial}{\partial p_\alpha}
\left( \kappa_{\alpha \beta} p_\beta
  - \frac{1}{\tau} p_\alpha \right)
+ \frac{1}{\tau} 
\frac{\partial}{\partial p_\alpha}
\frac{\partial}{\partial p_\alpha}
\\
& = &
- \frac{\partial}{\partial \vec{p}}
\cdot
\left( \tensor{\kappa} \cdot \vec{p}
  - \frac{1}{\tau} \vec{p} \right)
+ \frac{1}{\tau} \frac{\partial}{\partial \vec{p}}
\cdot \frac{\partial}{\partial \vec{p}} .
\label{eq:FPoperatorExplicit}
\eea
In order to take into account the fact that the probability
density is advected with the flow, we write the Fokker-Planck
equation as
\be
\label{eq:ConvectedFPEexplicit}
D_t P = {\cal L} P .
\ee
For the thermal average of an observable $A$ we thus have the
equation of motion
\be
D_t \left< A \right> = \left< {\cal L}^\dag A \right> ,
\ee
where ${\cal L}^\dag$ is the adjoint Fokker-Planck operator,
explicitly given as
\be
{\cal L}^\dag =
\left( \tensor{\kappa} \cdot \vec{p}
  - \frac{1}{\tau} \vec{p} \right)
\cdot
\frac{\partial}{\partial \vec{p}}
+ \frac{1}{\tau} \frac{\partial}{\partial \vec{p}}
\cdot \frac{\partial}{\partial \vec{p}} .
\ee
In particular, if we choose $\vec{p}$ as the observable, then
we find
\be
{\cal L}^\dag \vec{p} =
\tensor{\kappa} \cdot \vec{p} -
\frac{1}{\tau} \vec{p}
\ee
and hence
\be
\label{eq:MotionThermalAvPAlpha}
D_t \left< \vec{p} \right> =
\tensor{\kappa} \cdot \left< \vec{p} \right> -
\frac{1}{\tau} \left< \vec{p} \right> .
\ee
Similarly, one finds after a few lines of straightforward
algebra
\be
{\cal L}^\dag p_\alpha p_\beta =
p_\alpha p_\gamma \kappa_{\beta \gamma} +
p_\beta p_\gamma \kappa_{\alpha \gamma} -
\frac{2}{\tau}
\left( p_\alpha p_\beta - \delta_{\alpha \beta}
\right) ,
\ee
which implies
\bea
\nonumber
&&
D_t \left< p_\alpha p_\beta \right> -
  \left< p_\alpha p_\gamma \right> \kappa_{\beta \gamma} -
  \left< p_\beta p_\gamma \right> \kappa_{\alpha \gamma}
\\
& = &
- \frac{2}{\tau} \left(
\left< p_\alpha p_\beta \right> - \delta_{\alpha \beta}
\right)
\eea
or
\be
\label{eq:MotionThermalAvPAlphaPbeta}
\delta_t \left< p_\alpha p_\beta \right> =
  - \frac{2}{\tau} \left(
    \left< p_\alpha p_\beta \right> - \delta_{\alpha \beta}
      \right) .
\ee
Furthermore, from Eq.~\ref{eq:MotionThermalAvPAlpha} we find
\be
\label{eq:MotionTensorFrom1stMoments}
\delta_t \left(
\left< p_\alpha \right> \left< p_\beta \right>
\right) =
- \frac{2}{\tau}
\left< p_\alpha \right> \left< p_\beta \right> .
\ee
The last two equations are nothing but the constitutive equations of
the standard and the modified Oldroyd-B model,
respectively. Obviously, we have to set $\tau = 2$ in our reduced unit
system. The two equations differ by the unit tensor on the RHS. By
taking the difference between the two equations, one sees that this
term is directly related to thermal fluctuations.

Finally, we may now define, for $\alpha \ge 0$,
\be
\label{eq:tensorCFromLinearCombination}
\tensor{C} = \alpha \left< \vec{p} \vec{p}^T \right>
+ (1 - \alpha) \left< \vec{p} \right> \left< \vec{p}^T \right> .
\ee
From the previous equations of motion we then immediately find
(recall $\tau = 2$)
\be
\delta_t \tensor{C} = - \tensor{C} + \alpha \tensor{1} ,
\ee
i.~e. the equation of motion of the generalized Oldroyd-B model, which
thus turns out to be a linear combination of the standard and the
modified version. If we restrict the range of $\alpha$ to
$0 \le \alpha \le 1$, which seems reasonable, we actually have
a convex combination.

\section{The generalized Oldroyd-B model in the framework of the
  Navier-Stokes-Fokker-Planck system}
\label{sec:NSFP}

We have already seen that the Oldroyd-B model can be shown to
be dissipative, and this is true for both the standard version as well
as the modified version. It is also true for a suitable linear
combination thereof, i.~e. the generalized Oldroyd-B model. Similarly,
we have seen that both variants can be derived from the same
Navier-Stokes-Fokker-Planck (NSFP) system. The only difference is
the prescription how to obtain the macroscopic conformation tensor
entering the momentum equation from the moments of the Fokker-Planck
propagator $P$. 

In this section we wish to demonstrate that this underlying
NSFP system is dissipative as well, which should hardly be surprising,
in view of the results of the previous sections.

Let us therefore repeat the equations of motion of the NSFP system,
again using dimensionless units (cf. Eqs.~\ref{eq:OldroydB4},
\ref{eq:OldroydB5}, \ref{eq:ConvectedFPEexplicit}, 
\ref{eq:FPoperatorExplicit}):
\bea
\nabla \cdot \vec{v} & = & 0 , \\
D_t \vec{v} + \nabla p - \Gamma \nabla \cdot \tensor{C}
& = & \nabla^2 \vec{v} , \\
D_t P(\vec{r}, \vec{p}, t)
& = & {\cal L} P(\vec{r}, \vec{p}, t) ,
\eea
\be
{\cal L} =
- \frac{\partial}{\partial \vec{p}}
\cdot
\left( \tensor{\kappa} \cdot \vec{p}
  - \frac{1}{2} \vec{p} \right)
+ \frac{1}{2} \frac{\partial}{\partial \vec{p}}
\cdot \frac{\partial}{\partial \vec{p}} .
\ee
The field degrees of freedom for this system are the velocity flow field
$\vec{v}$ and the propagator $P$. In order to turn this into a closed
system, we still need to add a constitutive equation, which is here
the prescription how to calculate $\tensor{C}$ from $P$. Within
the investigations of the present paper, we of course should take
the prescription for the generalized Oldroyd-B model, i.~e.
Eq.~\ref{eq:tensorCFromLinearCombination}.

\subsection{Analysis of conservative dynamics}
\label{sec:conservativeNSFP}

For the conservative part of the NSFP system, we have the equations of motion
\bea
\nabla \cdot \vec{v} & = & 0 , \\
D_t \vec{v} + \nabla p - \Gamma \nabla \cdot \tensor{C}
& = & 0 , \\
D_t P(\vec{r}, \vec{p}, t)
& = & - \frac{\partial}{\partial \vec{p}}
\cdot \tensor{\kappa} \cdot \vec{p} \, P(\vec{r}, \vec{p}, t).
\eea
For being conservative, the dynamics must also conserve the underlying
Hamiltonian $\cal {H}$, which should be interpreted as the Helmholtz free 
energy of the system. 

We now assume that the free energy for this system may be written
as (in $d$ spatial dimensions)
\bea
\label{eq:HamiltonianNSFP}
{\cal H}_{\rm{NSFP}} & = & {\cal H}_1 + {\cal H}_2 + {\cal H}_3 + {\cal H}_4 ,
\\
{\cal H}_1 & = & \frac{1}{2} \int d \vec{r} \, \vec{v}^2 ,
\\
{\cal H}_2 & = & \frac{\Gamma}{2} (1 - \alpha)
\int d \vec{r} \, \left< \vec{p} \right>^2 ,
\\
{\cal H}_3 & = & \frac{\Gamma}{2} \alpha
\int d \vec{r} \, \left< \vec{p}^2 \right> ,
\\
{\cal H}_4 & = & \int d \vec{r} \int d \vec{p} \,\psi \left(P\right) ,
\eea
where $\psi$ is a scalar function of the propagator $P$. $\psi(P)$
must be chosen in such a way that the Hamiltonian is conserved, i.~e.
\be
\frac{d {\cal H}}{dt} = 0 .
\ee
We notice that in view of our constitutive equation we may write
\be
{\cal H}_2 + {\cal H}_3 =
\frac{\Gamma}{2} \int d \vec{r} \, \, \rm{tr} \tensor{C} .
\ee
From Secs. \ref{sec:ConservativeDynamics} and
\ref{sec:DissipativeDynamics} we recall that the
Hamiltonian of the generalized Oldroyd-B (GOB) model can
therefore be written as
\be
{\cal H}_{\rm{GOB}} = {\cal H}_1 + {\cal H}_2 + {\cal H}_3
- \frac{\Gamma}{2} \alpha \int d \vec{r} \, \rm{tr} \ln \tensor{C} .
\ee
We already derived that $d {\cal H}_{\rm{GOB}} / dt = 0$, and we
also found that ${\cal H}_{\rm{GOB}}$ is conserved even if we
omit the $\rm{tr} \ln \tensor{C}$ term. Therefore, we may just
refer to the results of Secs. \ref{sec:ConservativeDynamics} and
\ref{sec:DissipativeDynamics} to immediately conclude
\be
\label{eq:ThreeHamiltonian}
\frac{d}{dt} \left( {\cal H}_1 + {\cal H}_2
+ {\cal H}_3 \right) = 0 .
\ee
For the dynamics of ${\cal H}_4$ we only need to study the
dynamics of $P$, which we may write as
\bea
\partial_t P & = & {\cal L}_c P ,
\\
{\cal L}_c & = & - v_\alpha \partial_\alpha
- \frac{\partial}{\partial p_\alpha} \kappa_{\alpha \beta} p_\beta .  
\eea
For the operator ${\cal L}_c$ we note that incompressibility
implies the operator identities
\bea
v_\alpha \partial_\alpha & = & \partial_\alpha v_\alpha ,
\\
\frac{\partial}{\partial p_\alpha} \kappa_{\alpha \beta} p_\beta
& = &
\kappa_{\alpha \beta} p_\beta \frac{\partial}{\partial p_\alpha} .
\eea
With this it is straightforward to show, via integration by
parts, that ${\cal L}_c$ is skew-adjoint:
\be
\int d \vec r \int d \vec p \, f {\cal L}_c g =
- \int d \vec r \int d \vec p \, g {\cal L}_c f .
\ee
Furthermore, ${\cal L}_c$ satisfies a standard product rule:
\be
{\cal L}_c (f g) = f {\cal L}_c g + g {\cal L}_c f ,
\ee
and of course it also satisfies the chain rule. Now,

\bea
\nonumber
&&
\frac{d}{dt} {\cal H}_4
\\
\nonumber
& = &
\int d \vec r \int d \vec p \, \frac{\partial \psi}{\partial P}
{\cal L}_c P
\\
\nonumber
& = &
- \int d \vec r \int d \vec p \,
\left(
v_\alpha \frac{\partial \psi}{\partial P} \partial_\alpha P
+ \kappa_{\alpha \beta} p_\beta \frac{\partial \psi}{\partial P}
\frac{\partial P}{\partial p_\alpha}
\right)
\\
\nonumber
& = &
- \int d \vec r \int d \vec p \,
\left(
v_\alpha \partial_\alpha \psi
+ \kappa_{\alpha \beta} p_\beta
\frac{\partial \psi}{\partial p_\alpha}
\right)
\\
\nonumber
& = &
- \int d \vec r \int d \vec p \,
\left(
\partial_\alpha v_\alpha \psi
+ \frac{\partial}{\partial p_\alpha}
\kappa_{\alpha \beta} p_\beta \psi
\right)
\\
& = &
0 .
\eea
In other words, any differentiable function $\psi(P)$ will give
rise to a Hamiltonian ${\cal H}_4$ that is conserved.

Note that this analysis has relied heavily on the properties
of ${\cal L}_c$, and, in particular, the incompressibility
of the flow. Let us now assume a more general operator ${\cal L}_c$,
of which we only know that it is a first-order differential
operator that satisfies product rule and chain rule. Denoting
the adjoint operator with ${\cal L}_c^\dag$ (which is of course
also a first-order differential operator satisfying product
rule and chain rule), we may then proceed as follows:
\bea
\nonumber
\frac{d}{dt} {\cal H}_4
& = &
\int d \vec r \int d \vec p \, \frac{\partial \psi}{\partial P}
{\cal L}_c P
\\
\nonumber
& = &
\int d \vec r \int d \vec p \, P
{\cal L}_c^\dag \frac{\partial \psi}{\partial P}
\\
\nonumber
& = &
\int d \vec r \int d \vec p \, P 
\frac{\partial^2 \psi}{\partial P^2} {\cal L}_c^\dag P
\\
\nonumber
& = &
\int d \vec r \int d \vec p \, P
{\cal L}_c \left( P \frac{\partial^2 \psi}{\partial P^2}
\right)
\\
& = &
\int d \vec r \int d \vec p \, P \frac{\partial}{\partial P}
\left( P \frac{\partial^2 \psi}{\partial P^2} \right)
{\cal L}_c P .
\eea
This offers various possibilities to achieve $d {\cal H}_4 / dt = 0$:
Firstly, we may assume $\partial \psi / \partial P = 0$, which
would imply a constant Hamiltonian, which may as well be set to
zero. In other words, this would simply mean to discard
${\cal H}_4$ altogether. The second possibility is
$\partial^2 \psi / \partial P^2 = 0$, which however would give
rise to a function $\psi$ that varies linearly with $P$, or,
in other words, to a Hamiltonian that is not bounded from below.
This possibility must therefore be dismissed. Therefore, the
simplest non-trivial solution is provided by the condition
\be
P \frac{\partial^2 \psi}{\partial P^2} = A ,
\ee
where $A$ is some constant. The solution of this differential
equation is
\be \label{eq:functional_f}
\psi = A\, P \ln {P} ,
\ee
i.~e. a Boltzmann-like function, which for $A \ge 0$ is bounded from
below. Here we have ignored a linear and a constant term, by setting
the corresponding integration constants to zero.

\subsection{Analysis of dissipative dynamics}

The dissipative part of the equations of motion can be written as
\bea
\nabla \cdot \vec{v} & = & 0 , \\
\label{eq:DissipativeStokesNSFP}
\partial_t \vec{v} & = & \nabla^2 \vec{v} , \\
\label{eq:CandidateNSFP}
\partial_t P & = & \frac{1}{2}
\left( \frac{\partial}{\partial \vec{p}} \cdot \vec{p}
+ \frac{\partial^2}{\partial \vec{p}^2} \right) P =:
{\cal L}_d P .
\eea
For the dissipative dynamics of ${\cal H}_1$ we may directly refer
to the results of Sec.~\ref{sec:DissipativeDynamics}, where we showed
that $d {\cal H}_1 / dt \le 0$ as a result of viscous dissipation.

For ${\cal H}_2$ we first notice that the dissipative part of the
equation of motion for $\left< \vec{p} \right>$ is
(cf. Eq.~\ref{eq:MotionThermalAvPAlpha})
\be
\partial_t \left< \vec{p} \right> =
- \frac{1}{2} \left< \vec{p} \right> ,
\ee
resulting in
\be
\partial_t \left< \vec{p} \right>^2 =
- \left< \vec{p} \right>^2
\ee
or
\be
\label{eq:2ndHamiltonianNSFP}
\frac{d{\cal H}_2}{dt} =
- \frac{\Gamma}{2} (1 - \alpha)
\int d \vec{r}\, \left< \vec{p} \right>^2 .
\ee
At this point, it becomes clear that indeed we should restrict
the range of $\alpha$ to the interval $0 \le \alpha \le 1$, such
that indeed the conformation tensor is a \emph{convex} combination
of the second- and first-moment based expressions. If this condition
is satisfied, then indeed $d {\cal H}_2 / dt \le 0$.

For ${\cal H}_3$, we consider the dissipative part of
Eq.~\ref{eq:MotionThermalAvPAlphaPbeta},
\be
\partial_t \left< p_\alpha p_\beta \right> =
- \left< p_\alpha p_\beta \right> + \delta_{\alpha \beta} ,
\ee
from which we conclude
\be
\partial_t \left< \vec{p}^2 \right> =
- \left< \vec{p}^2 \right> + d 
\ee
(recall $d$ denotes the spatial dimension). Therefore
\be
\frac{d {\cal H}_3}{dt} = - \frac{\Gamma}{2} \alpha
\int d \vec{r} \, \left< \vec{p}^2 - d \right>  .
\ee
For the dynamics of ${\cal H}_4$ we study the properties of the
dissipative part of the Fokker-Planck operator, ${\cal L}_d$
(cf. Eq.~\ref{eq:CandidateNSFP}). Via integration by parts
it is easily shown that its adjoint operator is given by
\be
{\cal L}_d^\dag = \frac{1}{2} \left(
\frac{\partial^2}{\partial \vec{p}^2} -
\vec{p} \cdot \frac{\partial}{\partial \vec{p}} \right) .
\ee
Assuming
\be
{\cal H}_4 = A \int d \vec r \int d \vec p \, P \ln P ,
\ee
we thus find
\bea
\nonumber
\frac{d {\cal H}_4}{dt}
& = &
A \int d \vec r \int d \vec p \, (\ln P + 1) {\cal L}_d P
\\
\nonumber
& = &
A \int d \vec r \int d \vec p \, P {\cal L}_d^\dag (\ln P + 1)
\\
\nonumber
& = &
A \int d \vec r \left< {\cal L}_d^\dag (\ln P + 1) \right> .
\eea
Straightforward evaluation, combined with some regrouping
of terms, yields
\bea
\nonumber
&&
{\cal L}_D^\dag (\ln P + 1)
\\
& = &
\frac{1}{2} \left[
  \frac{1}{P} \frac{\partial^2 P}{\partial \vec{p}^2}
  + \vec{p}^2
  + \vec{p} \cdot \frac{\partial}{\partial \vec{p}} \ln P
  - \left( \vec{p} + \frac{\partial}{\partial \vec{p}} \ln P
  \right)^2 \right] .
\eea
Now,
\be
\left< \frac{1}{P} \frac{\partial^2 P}{\partial \vec{p}^2} \right>
=
\int d \vec p \, \frac{\partial^2 P}{\partial \vec{p}^2} = 0 ,
\ee
\be
\left< \vec{p} \cdot \frac{\partial}{\partial \vec{p}} \ln P \right>
=
\int d \vec p \, \vec{p} \cdot \frac{\partial P}{\partial \vec{p}}
= - d ,
\ee
such that
\be
\left< {\cal L}_D^\dag (\ln P + 1) \right>
=
\frac{1}{2} \left< \vec{p}^2 - d -
\left( \vec{p} + \frac{\partial}{\partial \vec{p}} \ln P
\right)^2 \right> .
\ee
Therefore, if we set $A = \Gamma \alpha$, we can combine the
results for ${\cal H}_3$ and ${\cal H}_4$ to yield
\be\label{eq:3rdAnd4thHamiltonianNSFP}
\frac{d\left({\cal H}_3 + {\cal H}_4\right)}{dt} =
- \frac{\Gamma}{2} \alpha \int d \vec{r} \,
\left< \left(
\vec{p} + \frac{\partial}{\partial \vec{p}} \ln {P}
\right)^2\right> \le 0 .
\ee
Therefore, the dynamics is indeed dissipative, and the Second Law
holds.

\begin{table*}[t]

  \begin{tabular}{| l | c | c |}
    \hline
    version & standard & modified \\
    \hline
    \hline
    momentum equation & \multicolumn{2}{c |}{
      $ D_t \vec{v} = - \nabla p + \Gamma \nabla \cdot \tensor{C}
      + \nabla^2 \vec{v} \quad\quad \nabla \cdot \vec{v} = 0 $
    } \\
    \hline
    constitutive relation &
    $ \delta_t \tensor{C} = - \tensor{C} + \tensor{1} $ &
    $ \delta_t \tensor{C} = - \tensor{C} $ \\
    \hline
    positivity of $\tensor{C}$ & strictly positive-definite &
    positive-semidefinite \\
    \hline
    Hamiltonian & \multicolumn{2}{c |}{
    $ {\cal H} = \int d \vec{r} \,
    \left( \vec{v}^2/2 + f (\tensor{C}) \right) $ } \\
    \hline
    free energy density &
    $ f = (\Gamma /2) \left( \textrm{tr} \tensor{C}
    - \textrm{tr} \ln \tensor{C} \right) $ &
    $ f = (\Gamma /2) \textrm{tr} \tensor{C} $ \\
    \hline
    driving force &
    $ \tensor{\chi} = (\Gamma / 2) \left(
    \tensor{1} - \tensor{C}^{-1} \right) $ &
    $ \tensor{\chi} = (\Gamma / 2) \tensor{1} $ \\
    \hline
    dissipation rate & \multicolumn{2}{c |}{
      $ \partial_t f = - (2/\Gamma) \textrm{tr}
      \left( \tensor{\chi} \cdot \tensor{C} \cdot
      \tensor{\chi} \right) $ } \\
    \hline
    $\tensor{C}$ from kinetic-theory moments &
    $C_{\alpha \beta} = \left< p_\alpha p_\beta \right>$ &
    $C_{\alpha \beta} = \left< p_\alpha \right>
    \left< p_\beta \right>$ \\
    \hline
    ensemble-defining quantity &
    conformation tensor & end-to-end vector \\
    \hline
  \end{tabular}
  \caption{Summary of results.}
  \label{tab:summary}

\end{table*}

\section{Summary, discussion and outlook}
\label{sec:Conclus}

Let us briefly summarize the main results of the present paper,
which are also presented in Table~\ref{tab:summary}.


We have discussed generalized Oldroyd-B models of the form
\bea
\nabla \cdot \vec{v} & = & 0 , \\
D_t \vec{v} + \nabla p - \Gamma \nabla \cdot \tensor{C}
& = & \nabla^2 \vec{v} , \\
\delta_t \tensor{C} & = & - \tensor{C} + \alpha \tensor{1} ,
\eea
with $0 \le \alpha \le 1$, where we focused on the extreme cases
$\alpha = 1$ (standard Oldroyd-B model) and $\alpha = 0$ (modified
Oldroyd-B model). We have seen that the dynamics conserves the
positivity of the conformation tensor, and that the model is
thermodynamically consistent for each choice of $\alpha$, where for
the free energy density we have to set
\be
f (\tensor{C}) =  \frac{\Gamma}{2} \left[ \textrm{tr} \tensor{C}
  - \alpha \textrm{tr} \ln \tensor{C} \right] .
\ee
We have also seen that all these models can be derived from the NSFP
system, where the standard model corresponds on a definition of the
conformation tensor based on the second moment of the propagator
(``tensorial theory''), while the modified model bases the definition
on the first moment (``vectorial theory''). The generalized model is
then simply the convex combination of the two extreme cases. The NSFP
system was shown to be dissipative for each choice of $\alpha$.

These results establish, from a formal point of view, that the
closure, which prescribes how to obtain the conformation tensor from
the NSFP system, is not unique but ambiguous. Importantly, we argued
that this is more than a formal mathematical exercise: Rather, the
ambiguity can be traced back to an ambiguity in the underlying
non-equilibrium ensemble, which in turn is a matter of \emph{choice}
(as a matter of fact: choice of the ensemble-defining slow variable).
Here it is important to realize that the conformation tensor that
enters the macroscopic momentum conservation equation is is a suitably
\emph{averaged} conformation tensor, and the details of the averaging
depend on the ensemble. The details of the closure then dictate both
the precise form of the resulting constitutive equation, and the
precise form of the underlying free energy. It seems to us that these
ensemble aspects, and their fairly far-reaching consequences, have up
to now not been fully appreciated, and are of significant importance
to the whole field of rheology. In the present study, we have
identified an ensemble based on the conformation tensor as the
underlying statistical-mechanical theory for the standard version of
the model, while the modified version builds on an ensemble based on
the end-to-end vector. Significantly, in the former case we have to
look at the limit $N \to \infty$, such that the constraining effect on
the single dumbbell is negligible, and thermal fluctuations are fully
present, while in the latter case we have to study $N = 1$, such that
thermal fluctuations are fully suppressed. In practice, this means
that in the modified case we base the theory not on the second but on
the first moment of the distribution.

As far as we know and understand, there is no fundamental and obvious
\emph{a priori} principle which would tell us that one of the two
versions is, in some sense, superior to the other one. A vague
intuitive feeling tells us that perhaps the version without
fluctuations might be more appropriate for the desired macroscopic
description, since, after all, the momentum conservation equation does
not include any fluctuations either. This is corroborated by the fact
that the derivation of
Ref.~\onlinecite{spillerSystematicDerivationHydrodynamic2021a}, which
was not based upon coupling the momentum conservation equation to a
Fokker-Planck system, but rather on direct coarse-graining combined
with GENERIC, also gives rise to the fluctuation-free version. We feel
that the best way to incorporate thermal fluctuations into rheological
models is to generalize the equations of motion to field-theoretic
stochastic differential equations, where the noise term represents
thermal fluctuations. The construction of such equations of motion for
viscoelastic models has already been worked out in
Ref.~\onlinecite{hutter_fluctuating_2018}, and we refer the interested
reader to that paper. Again, the GENERIC formalism provides a
straightforward guiding principle in this task. It is then reasonable
to assume that a model that describes fluctuations explicitly via
Langevin noise should \emph{not} include fluctuations in the
corresponding deterministic part. If this is correct, then this
consideration provides a strong argument in favor of the modified
model.

In this context, one should also realize that the modified model
(without noise) allows the trivial solution of a conformation tensor
(and polymer stress) that simply vanishes identically --- provided
this is compatible with the initial and boundary conditions. If this
latter condition holds, then the solution $\tensor{C} \equiv 0$ is
indeed expected to hold throughout the dynamics, regardless of
(possibly even turbulent) flow conditions. Note also that in the
special case of vanishing flow, where $\delta_t = \partial_t$, one may
solve the constitutive equation trivially, with the result of an
exponential relaxation towards zero (for the noise-free modified
model), or towards the unit tensor (standard model), which indicates
that in the modified model the state $\tensor{C} = 0$ is anything but
exotic. We believe that this should \emph{not} be dismissed as an
absurd unphysical result, which would invalidate the modified model.
Rather, we think it is likely that this actually describes the real
physics of the underlying microscopic model: If we turn off thermal
fluctuations completely, then it simply makes sense to assume that all
dumbbells will eventually shrink to zero extension, and will then
remain in that state, where they are also unable to modify the flow
and are rather transported as passive Langrangian particles. If this
is indeed true, then one should expect that the modified model is
able to provide non-trivial rheological behavior only for the version
with added Langevin noise.

We find one property of the modified model quite attractive: The
singularity of the Hamiltonian at $\tensor{C} = 0$, and the
corresponding singularity in the dissipation rate, are
removed. Therefore, the system might exhibit less dissipative
resistance against external driving, which might then perhaps lead to
an alleviated high Weissenberg number problem.

We thus see that these considerations lead to a number of unanswered
questions and speculations. It is therefore of high interest to test
the new model by numerical simulations, which is however left to
future work.

\begin{acknowledgments}
Stimulating discussions with J. Ravi Prakash are gratefully
acknowledged. Funded by the Deutsche Forschungsgemeinschaft (DFG,
German Research Foundation), Project No. 233630050-TRR 146.
\end{acknowledgments}

\section*{Author declarations}

The authors have no conflicts to disclose. All authors have
contributed equally to the preparation of this manuscript.

\section*{Data availability}

Data sharing is not applicable to this article as no new data were
created or analyzed in this study.

\appendix

\section{Proof of Eq. \ref{eq:ConstrainedAverageConfTensorGaussian}}
\label{sec:GaussianAverage}

We consider our system of $N$ dumbbells in thermal equilibrium, and
recall
\be
\hat{\vec{Q}} = N^{-1} \sum_i \vec{p}_i .
\ee
The probability distribution of $\hat{\vec{Q}}$ is Gaussian, which
(in $d$ spatial dimensions) can be written as
\be
\left< \delta \left( \hat{\vec{Q}} - \vec{Q} \right) \right> =
\left( \frac{N}{2 \pi} \right)^{d/2} \exp
\left( - \frac{N}{2} \vec{Q}^2 \right) .
\ee
We wish to calculate the constrained average of the conformation
tensor of, say, the first dumbbell, $\left[ p_{1 \alpha} p_{1 \beta}
  \right]$. Per definition we have
\be
\left[ p_{1 \alpha} p_{1 \beta} \right]
\left< \delta \left( \hat{\vec{Q}} - \vec{Q} \right) \right>
=
\left< \delta \left( \hat{\vec{Q}} - \vec{Q} \right)
p_{1 \alpha} p_{1 \beta} \right> .
\ee
The RHS may be evaluated by making use of the Fourier representation
of the delta function,
\bea
&&
(2 \pi)^d \delta \left( \hat{\vec{Q}} - \vec{Q} \right)
\\
\nonumber
& = &
\int d\vec{k} \exp( - i \vec{k} \cdot \vec{Q} )
\exp \left( i \frac{\vec{k}}{N} \cdot \vec{p}_1 \right) \ldots
\exp \left( i \frac{\vec{k}}{N} \cdot \vec{p}_N \right) .
\eea
Making use of the fact that the dumbbells are statistically
independent, and have identical properties, we find
\bea
\nonumber
&&
(2 \pi)^d \left< \delta \left( \hat{\vec{Q}} - \vec{Q} \right)
p_{1 \alpha} p_{1 \beta} \right>
\\
\nonumber
& = &
\int d\vec{k} \exp( - i \vec{k} \cdot \vec{Q} )
\left< \exp \left( i \frac{\vec{k}}{N} \cdot \vec{p}_1 \right)
p_{1 \alpha} p_{1 \beta} \right>
\\
& &
\left< \exp \left( i \frac{\vec{k}}{N} \cdot \vec{p}_2 \right)
\right>^{N - 1} .
\eea
Now,
\be
\left<
\exp \left( i \frac{\vec{k}}{N} \cdot \vec{p}_2 \right)
\right>
=
\exp \left( - \frac{\vec{k}^2}{2 N^2} \right)
\ee
and
\bea
\nonumber
&&
\left< \exp \left( i \frac{\vec{k}}{N} \cdot \vec{p}_1 \right)
p_{1 \alpha} p_{1 \beta} \right>
\\
\nonumber
& = &
- N^2
\frac{\partial}{\partial k_\alpha} \frac{\partial}{\partial k_\beta}
\left< \exp \left( i \frac{\vec{k}}{N} \cdot \vec{p}_1 \right) \right>
\\
\nonumber
& = &
- N^2
\frac{\partial}{\partial k_\alpha} \frac{\partial}{\partial k_\beta}
\exp \left( - \frac{\vec{k}^2}{2 N^2} \right)
\\
& = &
\left( \delta_{\alpha \beta} - \frac{k_\alpha k_\beta}{N^2} \right)
\exp \left( - \frac{\vec{k}^2}{2 N^2} \right) ,
\eea
resulting in
\bea
&&
(2 \pi)^d \left< \delta \left( \hat{\vec{Q}} - \vec{Q} \right)
p_{1 \alpha} p_{1 \beta} \right>
\\
\nonumber
& = &
\int d\vec{k} \exp( - i \vec{k} \cdot \vec{Q} )
\left( \delta_{\alpha \beta} - \frac{k_\alpha k_\beta}{N^2} \right)
\exp \left( - \frac{\vec{k}^2}{2 N} \right)
\\
\nonumber
& = &
\left( \delta_{\alpha \beta} + N^{-2} \frac{\partial}{\partial Q_\alpha}
\frac{\partial}{\partial Q_\beta} \right)
\int d\vec{k} \exp( - i \vec{k} \cdot \vec{Q} )
\exp \left( - \frac{\vec{k}^2}{2 N} \right)
\\
\nonumber
& = &
\left( \delta_{\alpha \beta} + N^{-2} \frac{\partial}{\partial Q_\alpha}
\frac{\partial}{\partial Q_\beta} \right)
\left( 2 \pi N \right)^{d/2}
\exp \left( - \frac{N}{2} \vec{Q}^2 \right) .
\eea
Taken together, one thus finds
\bea
\nonumber
&&
\left[ p_{1 \alpha} p_{1 \beta} \right]
\\
\nonumber
& = &
\exp \left( + \frac{N}{2} \vec{Q}^2 \right)
\left( \delta_{\alpha \beta} + N^{-2} \frac{\partial}{\partial Q_\alpha}
\frac{\partial}{\partial Q_\beta} \right)
\exp \left( - \frac{N}{2} \vec{Q}^2 \right)
\\
& = &
Q_\alpha Q_\beta + \left( 1 - N^{-1} \right) \delta_{\alpha \beta} .
\eea


\end{document}